\documentclass[aj]{emulateapj}
\usepackage{natbib}
\usepackage{rotating}
\usepackage{threeparttable}
\usepackage{color}
\usepackage{graphicx}
\usepackage{multirow}

\slugcomment{To appear in AJ}

\shorttitle{Faraday structure challenge}

{\date\today}

\begin{document}
\title{Comparison of algorithms for determination of rotation measure and Faraday structure \,\,I. 1100 - 1400 MHz}

\author{X. H. Sun\altaffilmark{1}, L. Rudnick\altaffilmark{2}, 
Takuya Akahori\altaffilmark{1},
C. S. Anderson\altaffilmark{1},
M. R. Bell\altaffilmark{3}, 
J. D. Bray\altaffilmark{4},
J. S. Farnes\altaffilmark{1}, 
S. Ideguchi\altaffilmark{5},
K. Kumazaki\altaffilmark{5},
T. O'Brien\altaffilmark{2},
S. P. O'Sullivan\altaffilmark{1},
A. M. M. Scaife\altaffilmark{4},
R. Stepanov\altaffilmark{6,7}, 
J. Stil\altaffilmark{8},
K. Takahashi\altaffilmark{9},
R. J. van Weeren\altaffilmark{10},
M. Wolleben\altaffilmark{8}}

\altaffiltext{1}{Sydney Institute for Astronomy, School of Physics, The University of Sydney, NSW 2006, Australia; x.sun@physics.usyd.edu.au} 

\altaffiltext{2}{Minnesota Institute for Astrophysics, School of Physics and Astronomy, University of Minnesota, 116 Church Street SE, Minneapolis, MN  55455, USA; larry@umn.edu}

\altaffiltext{3}{Max Planck Institute for Astrophysics, Karl-Schwarzschild-Str. 1, 85748 Garching, Germany}

\altaffiltext{4}{Department of Physics \& Astronomy, University of Southampton, Highfield, Southampton SO17 1BJ, UK}

\altaffiltext{5}{University of Nagoya, Furo-cho, Chikusa-ku, Nagoya 464-8601, Japan}

\altaffiltext{6}{Institute of Continuous Media Mechanics, Korolyov str. 1, 614061 Perm, Russia}

\altaffiltext{7}{Perm National Research Polytechnic University, Komsomolskii Av. 29, 614990 Perm, Russia}

\altaffiltext{8}{Department of Physics and Astronomy, University of Calgary, 2500 University Drive NW, Calgary AB T2N 1N4, Canada}

\altaffiltext{9}{University of Kumamoto, 2-39-1, Kurokami, Kumamoto 860-8555, Japan}

\altaffiltext{10}{Harvard-Smithsonian Center for Astrophysics, 60 Garden Street, Cambridge, MA 02138, USA}

\begin{abstract} 
Faraday rotation measures (RMs) and more general Faraday structures are key 
parameters for studying cosmic magnetism and also are sensitive probes of faint 
ionized thermal gas. There is a need to define what derived quantities are 
required for 
various scientific studies, and then to address the challenges in determining 
Faraday structures. A wide variety of algorithms have been proposed to 
reconstruct these structures. In preparation for the Polarization Sky Survey of the Universe's 
Magnetism (POSSUM) to be conducted with the Australian Square Kilometre Array 
Pathfinder (ASKAP) and the ongoing Galactic Arecibo L-band Feeds Array 
Continuum Transit Survey (GALFACTS), we run a Faraday structure determination 
data challenge to benchmark the currently available algorithms 
including Faraday synthesis (previously called RM synthesis in the literature), 
wavelet, compressive sampling and $QU$-fitting. The input models include sources 
with one Faraday thin component, two Faraday thin components and one Faraday 
thick component. The frequency set is similar to POSSUM/GALFACTS with a 
300-MHz bandwidth from 1.1 to 1.4~GHz. We define three figures of merit 
motivated by the underlying science: a) an average RM weighted by polarized 
intensity, $\rm RM_{wtd}$, b) the separation $\Delta\phi$ of two Faraday 
components and c) the reduced chi-squared $\chi_r^2$. Based on the current test 
data of signal to noise ratio of about 32, we find that: (1) When only one 
Faraday thin component is present, most methods perform as expected, with 
occasional failures where two components are incorrectly found; (2) For two 
Faraday thin components, $QU$-fitting routines perform the best, with errors 
close to the theoretical ones for $\rm RM_{wtd}$, but with significantly higher 
errors for $\Delta\phi$. 
All other methods including standard Faraday synthesis 
frequently identify only one component when $\Delta\phi$ is below or near the width of the Faraday point spread function; 
(3) No methods, as currently implemented, work well for Faraday 
thick components due to the narrow bandwidth; (4) There exist combinations of two Faraday components which 
produce a large range of acceptable fits and hence large uncertainties in the derived single RMs; 
in these cases, different RMs lead to the same $Q,\,U$ behavior, so no method 
can recover a unique input model. 
Further exploration 
of all these issues is required before upcoming surveys will be able to provide 
reliable results on Faraday structures. 
\end{abstract}


\section{Introduction}

Cosmic magnetism is one of the key science projects for the future Square 
Kilometre Array (SKA), which will measure Faraday rotation measures (RMs) of 
 tens of millions of background radio sources to reveal how cosmic magnetic 
fields are generated and how they evolve over cosmic time \citep{gbf04}. It is 
therefore crucial to appropriately define and determine RMs, which can be very 
challenging when a radio source has more than a single Faraday component, which we will hereinafter designate as a composite Faraday structure.
A source can be identified as being a Faraday composite if its fractional linear polarization is not 
constant as a function of wavelength squared ($\lambda^2$). 

Faraday rotation occurs when a linearly polarized radio wave propagates in a 
magneto-ionic medium. In the simplest case, the polarization angle is rotated 
by an amount proportional to the wavelength squared, providing 
the definition of RM as 
\begin{equation}\label{eq:rmdef}
\chi(\lambda^2)=\chi_0+{\rm RM}\lambda^2,
\end{equation}
where $\chi(\lambda^2)$ is the polarization angle observed at $\lambda^2$ and 
$\chi_0$ is a constant. The RM in the simple case of a foreground screen is 
proportional to the integral of thermal electron density multiplied by magnetic 
field parallel to the line of sight from the source to the observer. A positive 
RM indicates a magnetic field pointing towards the observer. 

RMs of background {\em compact} extragalactic radio sources have been used to 
study magnetic fields in Galactic objects such as H\,{\scriptsize II} regions 
\citep{hmg11}, high velocity clouds \citep{hmb+13}, and supernova remnants 
\citep{srw+11}. On larger scales, they are used to probe magnetic fields in the 
Galaxy \citep{hmbb97,sr10,wfl+10,ptkn11,vbs+11,jf12,arkg13}, 
in nearby 
\citep{hbb98, ghs+05, mmg+12} and in high redshift galaxies \citep{bml13}. 
On yet larger 
scales, RM probes have been used for galaxy clusters including embedded radio 
galaxies \citep{bvb+13} and are proposed for cosmic webs 
\citep{ar10,ar11,aktr14}. 

Most of the currently available RMs were obtained by 
linearly fitting polarization angles versus $\lambda^2$ over a narrow range. 
For example, the catalog by \citet{tss09} containing 37\,543 RMs was based on 
polarization angles at 1364.9~MHz and 1435.1~MHz from the NRAO VLA Sky Survey 
\citep{ccg+98}. However, the linear relation between polarization angle and 
$\lambda^2$ is violated when a radio source has multiple Faraday components 
or Faraday depolarization effects occur, 
and RMs from the linear fit of polarization angle versus $\lambda^2$ are not 
reliable \citep{frb11,sbr+12}. Although the influence of an individual RM error 
for studying foreground magnetic fields can be largely eliminated with a sufficiently dense 
RM grid, variances in RMs can still remain. These variances are hard to 
interpret and can confuse measurements of small scale Faraday structure, as 
well as the information on the sources' intrinsic Faraday structures. A more 
sophisticated understanding of RMs and development of algorithms to determine 
them are certainly necessary. 

RMs of {\em diffuse} polarized emission have also been used to study the 
magnetic field in supernova remnants \citep{kb09,hgk+10,srw+11}, 
in the Galaxy 
\citep{hkd03b,skd09} and nearby galaxies \citep{bhb10, fbs+11,mmg+12}. 
The RM structure of 
the radio galaxy Centaurus A indicates the existence of faint ionized thermal 
gas in its radio lobes \citep{sfm+13}. Specific Faraday signatures can manifest 
the helical structure of the magnetic field \citep{bs14,hf14}. The diffuse emission 
is usually modeled as a slab with a uniform mixture of synchrotron-emitting and 
Faraday-rotating plasmas \citep{bur66}. In this case, not only the RM but also 
the Faraday extent of a slab are essential to properly understand the source 
structures. A single RM derived by fitting polarization angles versus 
$\lambda^2$ is unable to represent the distribution of RMs within the slab 
\citep{sbs+98}. To recover the complete Faraday structure of the diffuse 
emission, more advanced methods are required.  

Recently, there has been considerable progress in Faraday structure 
determination capabilities. Many observational facilities such as the 
Very Large Array (VLA), the Australian Telescope Compact Array 
(ATCA), the LOw Frequency ARray (LOFAR), the Murchison Widefield Array (MWA), 
and the Giant Metrewave Radio Telescope (GMRT) are now equipped with 
multi-channel broadband polarimeters which significantly reduces the 
ambiguities in characterizing source polarization properties. 

There are large-scale surveys underway such as the Galactic Magneto-Ionic 
Medium Survey \citep[GMIMS,][]{wlh+10} and the Galactic Arecibo L-band Feeds 
Array Continuum Transit Survey \citep[GALFACTS,][]{ts10}, and surveys to be 
conducted before the full SKA such as the Polarization Sky Survey of the 
Universe's Magnetism \citep[POSSUM,][]{glt10} with the Australian SKA 
Pathfinder (ASKAP). These surveys aim to advance our understanding on magnetism 
by dissecting the Faraday structure of the magneto-ionic medium. It is 
therefore crucial to decide which algorithms should be applied to the dataset 
to produce a reliable interpretation. 

Motivated by the need for competitive algorithms to reconstruct Faraday 
structures and the advent of many methods, we initiated a data challenge aiming 
to benchmark all the current methods. In this first step, we focus on the POSSUM and 
GALFACTS configurations, and use a similar frequency range covering 300~MHz 
from 1.1 to 1.4~GHz. The band is split into $300\times1$-MHz channels. This 
paper is organized as follows. We introduce the algorithms which were included 
in the 
data challenge in Sect.~\ref{methods}, describe the construction of the suite 
of tests in Sect.~\ref{data}, present the benchmark results in 
Sect.~\ref{results},  discussions in Sect.~\ref{discussions}, and conclusions 
in Sect.~\ref{conclusions}.

\section {Algorithms}\label{methods}

Following \citet{bur66}, the observed complex-valued polarized intensity at 
$\lambda^2$, $P(\lambda^2)$, can be written as  
\begin{equation}\label{eq:PF}
P(\lambda^2)=Q(\lambda^2)+iU(\lambda^2)=\int_{-\infty}^\infty F(\phi)\exp\left(2i\phi\lambda^2\right){\rm d}\phi. 
\end{equation} 
Here $\phi$, the Faraday depth, is defined as
\begin{equation}\label{eq:defF}
\phi = K\int_{\vec{r}}^{0} n_e\vec{B}\cdot\vec{{\rm d}l}, 
\end{equation}
where $K$ is a constant, $\vec{r}$ is the position of emission inside a source, 
$n_e$ is the electron density, $\vec{B}$ is the magnetic field and 
$\vec{{\rm d} l}$ is the distance increment along the line of sight, and 
$F(\phi)$, the Faraday spectrum 
(previously called Faraday dispersion function), represents the complex-valued 
polarized intensity at $\phi$. For a simple source that has polarized emission 
only at $\phi_0$, the Faraday spectrum is $F(\phi)=F_0\delta(\phi-\phi_0)$, 
where $F_0=f_0\exp(2i\chi_0)$ with $f_0$ being the polarized intensity at zero 
wavelength. The observed complex-valued polarized intensity is thus 
$P(\lambda^2)=f_0\exp\left[2i(\chi_0+\phi_0\lambda^2)\right]$, and in this 
simple case  RM=$\phi_0$. $F(\phi)$ is the key quantity that many techniques 
aim to extract from observations, in order to study the structure of the 
magneto-ionic medium. 

\begin{table*}
\begin{center}
\caption{Nomenclature.\label{tbl:nomen}}
\begin{tabular}{llp{5cm}}
\hline\hline
  Term &  Previous names & Description \\
\hline
Rotation Measure     & & RM, defined by Equation~(\ref{eq:rmdef})\\
Faraday depth        & & $\phi$, defined by Equation~(\ref{eq:defF})\\
Faraday spectrum     & Faraday dispersion function\tablenotemark{a}& 
                       $F(\phi)$, the complex-valued linearly polarized flux as a 
                       function of $\phi$ \\
Faraday synthesis    & RM synthesis\tablenotemark{a} & The process of deriving 
                       $F(\phi)$ from $P(\lambda^2)$ with a Fourier transform.\\
Faraday Point Spread Function & RM spread function\tablenotemark{b} & 
                                FPSF, the Fourier transform of the weighting 
                                function in the $\lambda^2$ domain.\\
Faraday clean       & RM clean\tablenotemark{b} & The process of deconvolving 
                      an observed $F(\phi)$ into an input one using the 
                      ``clean'' technique\\
3D Faraday synthesis & Faraday synthesis\tablenotemark{c} & Performing a 3D 
                       Fourier transform of complex-valued visibility data from the 
                       (u,v,$\lambda^2$) space to the (x,y,$\phi$) space\\
\hline
\end{tabular}
\tablecomments{$\rm^a$ \citet{bd05}; $\rm^b$ \citet{hea09}; $\rm^c$ \citet{be12}}
\end{center}
\end{table*}

The methods that are included in the data challenge either a) are open-ended, 
i.e. decomposing the Faraday spectra into some basis functions, and include 
Faraday synthesis \citep[previously called RM synthesis,][]{bd05,be12,kai+14} 
with Faraday clean \citep[previously called RM clean,][]{hea09}, wavelet 
\citep{fssb10,fssb11} and compressive sampling \citep{lbch11,ast12} or b) 
assume models of the synchrotron-emitting and Faraday rotating medium to fit 
the observed $P(\lambda^2)$, {\it viz.} $QU$-fitting \citep{frb11,sbr+12,ita+14}. When 
reporting the results many of the techniques involved a search by human eyes to 
determine whether some features in $F(\phi)$ are real. 
  
Note that there are various terms in the literature used to describe the 
polarization and Faraday properties of sources, and the processes and products 
of transforming from observations of the Stokes parameters as a function of 
wavelength into a Faraday depth space. Although these terms have historical 
value, they are becoming increasingly discordant with the actual applications, 
and are often misused. In this paper, and as a suggestion to the community 
moving forward, we adopt the terminology described in Table~\ref{tbl:nomen}.

\subsection{Open-ended methods}

\subsubsection{Faraday synthesis}
The complex-valued polarized intensity $P(\lambda^2)$ and the Faraday spectrum $F(\phi)$ are a 
Fourier transform pair from Eq.~(\ref{eq:PF}). Because, in 
practice, there are neither observations at $\lambda^2<0$ nor at all 
$\lambda^2>0$, the inverse Fourier transform of $P(\lambda^2)$, 
$\widetilde{F(\phi)}$, is actually a convolution of the true Faraday spectrum 
$F(\phi)$ and the Faraday Point Spread Function (FPSF, previously called RM 
spread function, see Table~\ref{tbl:nomen}) $R(\phi)$ which is the Fourier 
transform of the weighting function $W(\lambda^2)$. The weighting function 
depends on the frequency band coverage and weights of each of the individual 
frequency channels. \citet{bd05} have given a detailed description and also 
presented several formulae that can be easily applied to observations, as below
\begin{equation}\label{eq:rms}
\begin{array}{rcl}
\widetilde{F(\phi)}&\equiv& F(\phi)\star R(\phi) \\
&=& \mathcal{K} \sum W(\lambda_i^2)P(\lambda_i^2)\exp\left[-2i\phi(\lambda_i^2-\lambda_0^2)\right]\\[3mm]
R(\phi) &=& \mathcal{K}\sum W(\lambda_i^2)\exp\left[-2i\phi(\lambda_i^2-\lambda_0^2)\right]\\[3mm]
\lambda_0^2 &=& \mathcal{K}\sum W(\lambda_i^2)\lambda_i^2\\[3mm] 
\mathcal{K} &=& \left(\sum W(\lambda_i^2)\right)^{-1},
\end{array}
\end{equation}
where the sums are made over all the frequency channels, $\lambda_i^2$ is the 
wavelength squared of each channel, $\lambda_0^2$ is the weighted average of 
$\lambda_i^2$ and $\mathcal{K}$ is the normalization constant. For the current 
data challenge, the weighting function is $W(\lambda^2)=1$ for 
$\lambda_{\rm min}^2<\lambda^2<\lambda_{\rm max}^2$ and $W(\lambda^2)=0$ 
elsewhere. We refer to the methods employing the above formulae as the standard 
Faraday synthesis.

Three quantities can be used to characterize the limitations of recovering the 
Faraday spectrum from observations with a given frequency setup: the full width 
at half maximum (FWHM) of the FPSF, or equivalently, the resolution in Faraday 
depth space $\Delta\phi_{\rm FPSF}$, the maximal Faraday depth 
$\phi_{\rm max}$, and the largest Faraday depth scale to which the method is 
sensitive, $\delta\phi$. Following \citet{bd05}, these quantities are determined 
by the minimum wavelength squared $\lambda_{\rm min}^2$, the maximum wavelength 
squared $\lambda_{\rm max}^2$ and the interval of wavelength squared 
$\delta \lambda^2$ 
fixed by the channel width in frequency for the observations as 
\begin{eqnarray}
\Delta\phi_{\rm FPSF} & = & \frac{2\sqrt{3}}{\lambda_{\rm max}^2-\lambda_{\rm min}^2}\nonumber\\[3mm]
|\phi_{\rm max}| & = & \frac{\sqrt{3}}{\delta\lambda^2} \\[3mm]
\delta\phi & = & \frac{\pi}{\lambda_{\rm min}^2} \nonumber. 
\end{eqnarray}
Note that $\phi_{\rm max}$ refers to Faraday thin components and 
$\delta\phi$ refers to Faraday thick components. 
For the frequency setup of the test data, the values of 
$\lambda_{\rm min}^2$, $\lambda_{\rm max}^2$, $\delta\lambda^2$, 
$\Delta\phi_{\rm FPSF}$, $\phi_{\rm max}$, and $\delta\phi$ are listed in 
Table~\ref{tbl:keyparams}.

\begin{table}[!htbp]
\begin{center}
\caption{The wavelength range and several key parameters for the data 
challenge.\label{tbl:keyparams}}
\begin{tabular}{ll}
\hline\hline
$\lambda_{\rm min}^2$   &  $4.6\times10^{-2}$ m$^2$\\
$\lambda_{\rm max}^2$   &  $7.4\times10^{-2}$ m$^2$\\
$\delta\lambda^2$       &  $6.6\times10^{-5}$ m$^2$\\
$\Delta\phi_{\rm FPSF}$ &  122~rad~m$^{-2}$ \\
$|\phi_{\rm max}|$      &  26319~rad~m$^{-2}$ \\
$\delta\phi$            &   68~rad~m$^{-2}$\\
\hline
\end{tabular}
\end{center}
\end{table}

A deconvolution method similar to the ``clean-algorithm" widely used for radio 
synthesis images proposed by \citet{hog74} was developed by \citet{hea09}, 
which we refer to as the standard Faraday clean (Table~\ref{tbl:nomen}). The 
method first searches for the peak in $|\widetilde{F(\phi)}|$, and then 
subtracts the FPSF, $R(\phi)$, multiplied by the peak attenuated by a loop gain 
factor from $\widetilde{F(\phi)}$. The process iterates until the residual 
$|\widetilde{F(\phi)}|$ reaches a threshold or the number of loops exceeds a 
certain value. The clean components are then convolved to a Gaussian with a 
FWHM of $\Delta\phi_{\rm FPSF}$ and added back to the residuals to form the 
best estimate of $F(\phi)$. 

We denote a method based on Faraday synthesis and Faraday clean by ``FS-" 
followed by the abbreviation of the respective participant's name.\footnote{AS=Anna Scaife; JF=Jamie Farnes; JS=Jeroen Stil; KK=Kohei Kumazaki; LR=Lawrence Rudnick; MB=Michael Bell; MW=Maik Wolleben; RS=Rodion Stepanov; RvW=Reinout van Weeren; SO'S=Shane O'Sullivan; TO'B=Tim O'Brien and XS=Xiaohui Sun} 
Among the algorithms, FS-JF, FS-KK, FS-LR, and FS-RvW use a standard Faraday 
synthesis and Faraday clean described as above; FS-MW first bins the data to 
achieve uniform sampling in $\lambda^2$; FS-MB first regrids the data so that 
it is equally spaced in $\lambda^2$ and applies a Fast Fourier Transform to 
calculate $\widetilde{F(\phi)}$; FS-MBm is the same as FS-MB but with an extra 
step to determine the number of clean components with a maximum likelihood 
analysis \citep{boce13}.

\subsubsection{Wavelets}
The wavelet decomposition of the Faraday spectrum $F(\phi)$ into wavelet
coefficients $w(a, \phi)$ can be performed by the transformation of the
polarized intensity $P(\lambda^2)$ as suggested by \citet{fssb10}
\begin{equation}\label{eq:wavelet1}
w(a, \phi)=\frac{1}{\pi}\int_{-\infty}^{\infty}P(\lambda^2)\exp(-2i\phi\lambda^2)\hat{\psi}^*(-2a\lambda^2){\rm d}\lambda^2,
\end{equation}
where $\hat{\psi}$ is the Fourier transform of the analyzing wavelet, $a$
defines the scale, and $\phi$ is the Faraday depth of the wavelet center. The
Faraday spectrum $F(\phi)$ can be synthesized using the inverse transform
\begin{equation}\label{eq:wavelet2}
F(\phi)=\frac{1}{C_\psi}\int_{-\infty}^{\infty}\int_{-\infty}^{\infty}\psi\left(\frac{\phi-b}{a}\right) w(a,b)\frac{{\rm d}a{\rm d}b}{a^2}
\end{equation}
where $b$ is the position of the wavelets, and 
$C_\psi=1$ for the typically chosen wavelet, the so-called ``Mexican hat"
$\psi(\phi)=(1-\phi^2)\exp(-\phi^2/2)$.

The wavelet-based algorithm used here allows a combination of the Faraday
Synthesis procedure and the wavelet filtering. Coefficients $w(a,\phi)$ in
Eq.~(\ref{eq:wavelet2}) are not necessarily the same as those obtained from
Eq. (\ref{eq:wavelet1}), because there can be some additional constraints to
filter an extraneous signal, such as a noise, or to complement a signal using
specific assumptions \citep[e.g.][]{fssb11}. Another general purpose of the
wavelet application is to provide a multi-scale structure analysis.
Coefficients $w(a,\phi)$ represent a spectral composition of $F(\phi)$ locally
in Faraday space and its distribution over recognizable scales, set by the
wavelength range of a specific radio telescope \citep{bfss12}. The advantage of
wavelet decomposition can be considerable when the range of recognizable scales
in Faraday space is sufficiently wide, which means a large ratio of maximum to
minimum wavelengths, 
namely $(\lambda_{\rm max}/\lambda_{\rm min})^2\gg1$.
The current data challenge only allows for a maximum scale separation 
up to 1.6. This is not ideal, but is slightly above the principal scale 
resolution of the wavelets (typically estimated as 1.3). Therefore, wavelet analysis is possible, but cannot produce optimal results over the 1100 -- 1400 MHz band tested here. 
For a simple source, the wavelet method gives the same result as Faraday
synthesis \citep[e.g.][]{fssb11}. We refer to this method as Wavelet-RS.

\subsubsection{Compressive sampling}
In this method, it is assumed that the Faraday spectrum $F(\phi)$ can be 
represented by a sparse sample in a set of analysis functions, namely 
${\bf f}={\bf Z}{\bf \xi}$, where ${\bf f}=\{f_1, f_2,\ldots,f_M\}^T$ is the 
Faraday spectrum, ${\bf Z}$ is the $M\times M$ transform matrix for 
the representation, and ${\bf \xi}=\{\xi_1,\xi_2,\ldots,\xi_M\}^T$ is the 
coefficient. The reconstruction of the Faraday spectrum then turns 
to the minimization problem
\begin{equation}
{\rm min}\sum_{m=1}^{M}|\xi_m| \quad\quad {\rm subject\,to} \quad\quad 
|\mathbf{Yf}-\mathbf{P}|^2\leq(\beta\sigma)^2, 
\end{equation}
where $Y_{nm}=\exp(2i\phi_m\lambda^2_n)$ is the matrix translating the Faraday 
spectrum to the observed complex-valued polarized intensity 
${\bf P}=\{P_1,P_2,\ldots,P_N\}^T$, $\sigma$ is the noise in $Q$ and $U$ 
in each frequency channel, $\beta=\sqrt{N}$, $m=1,2,\ldots,M$ and 
$n=1,2,\ldots,N$, where $M$ is the number 
of sparse samples and $N$ is the number of frequency channels. 

The two methods used in the data challenge that are based on compressive 
sampling are: CS-XS developed by \citet{lbch11} using the Daubechies D8 wavelet 
transforms, and CS-JS devised by \citet{ast12} using a boxcar and 
$\delta$-function dictionary. Note that for some tests, only the 
$\delta$-function library was used and the decision of which function 
basis to use was made based on the growth plots 
(all the top row grayscale plots in \citealt{ast12}).

\subsection{Model fitting}\label{sect:qu-fitting}

The $QU$-fitting method selects a number of independent Faraday components to 
produce a net $Q(\lambda^2)$ and $U(\lambda^2)$ versus $\lambda^2$ followed by 
a minimization procedure. A detailed 
description of the physical models used in $QU$-fitting has been given by e.g. 
\citet{sbs+98}, \citet{frb11} and \citet{sbr+12}. Depending on the relative 
distribution of synchrotron-emitting and Faraday-rotating plasmas as well as 
the distribution of Faraday depths, the models that are used in the current 
data challenge can be categorized as follows. Note that composite structures 
based on $F(\phi)$ are also possible, but not included in the current tests.
\begin{itemize}

\item {\bf Simple Faraday screen}, where a synchrotron emitting region lies 
behind a Faraday rotating screen. The polarized intensity is then
\begin{equation}\label{eq:fs2}
P(\lambda^2)=p\exp\left[2i(\chi+\phi\lambda^2)\right]
\end{equation}
If there are a small number of separate synchrotron emitting regions, then 
\begin{equation}\label{eq:fs3}
P(\lambda^2)=p_1\exp\left[2i(\chi_1+\phi_1\lambda^2)\right]+p_2\exp\left[2i(\chi_2+\phi_2\lambda^2)\right]+\ldots
\end{equation}

\item {\bf Slab}, which represents a mixture of thermal and non-thermal 
plasmas. The synchrotron emissivity is uniform and the Faraday depth either 
increases or decreases linearly along the line of sight. The Faraday depth of 
the front edge of the screen is $\phi_0$ and the extent of the screen in Faraday depth is 
$\phi_s$. The expected polarized intensity can be written as 
\begin{equation}\label{eq:slab}
P(\lambda^2)=p_0\frac{\sin(\phi_s\lambda^2)}{\phi_s\lambda^2}\exp\left[2i(\chi_0+\phi_0\lambda^2+\frac{1}{2}\phi_s\lambda^2)\right]
\end{equation}  

\end{itemize}

The $QU$-fitting methods try to fit one or more of the above models to the 
observed $Q(\lambda^2)$ and $U(\lambda^2)$ versus $\lambda^2$ with various 
optimization schemes. The $QU$-fitting methods used for this data challenge 
are: QU-AS, using Markov Chain Monte Carlo with Metropolis-Hasting sampler and 
Bayesian evidence to distinguish between one-component and two-component 
models, as described by \citet{sh12}; QU-TO'B, using least-squared fits and 
QU-SO'S, using maximum likelihood. For QU-TO'B and QU-SO'S, standard Faraday 
synthesis and Faraday clean are run first to provide initial values for the 
parameters. To distinguish between different models, QU-SO'S implements a 
Bayesian information criterion to penalize models with more parameters, while 
QU-TO'B starts with the simplest models, and then adds more parameters only if 
the reduced chi-squared goes down.

\section{Construction of the data challenge}\label{data}

The data challenge is run for a suite of test data, where each test is a 
simulated data set containing 300 values of total intensity $I_\nu$, 
Stokes $Q_\nu$ and $U_\nu$ for each frequency $\nu$, where 
$\nu=1.100,\,1.101,\,1.102,\,\ldots,\,1.399$~GHz. The total intensity is 
arbitrary. For polarization, we derived $Q_\nu$ and $U_\nu$ from an 
input model of $F(\phi)$ according to Eq.~(\ref{eq:PF}), and then added 
Gaussian random noise with $\sigma_\nu=1$ to each of $Q_\nu$ and $U_\nu$. The 
rms noise for the full band is thus $\sigma=\sigma_\nu/\sqrt{N}$, where $N=300$ 
is the number of channels. The signal-to-noise ratio (SNR) was set to be 
$10^{1.5}\approx32$, which is well above the expected detection threshold of 
$8\sigma$ 
for POSSUM \citep{gsk12}. This ensures that all the methods described in 
Sect.~\ref{methods} should be able to easily detect the signals. In practice, 
$q_\nu=Q_\nu/I_\nu$ and $u_\nu=U_\nu/I_\nu$ were provided instead of $Q_\nu$ 
and $U_\nu$ as most of the algorithms could use them to remove the influence of 
any non-zero spectral indices which might complicate the Faraday 
clean process \citep[see][for a detailed discussion]{bd05}. 

Finally, 17 data sets were constructed for the benchmark test and they are 
listed in Table~\ref{tbl:all}. These data cover three types of input models for 
the Faraday spectrum as follows:
\begin{itemize}
\item {\it One Faraday thin component.} The Faraday spectrum is 
simply a $\delta$-function, $F(\phi)=F_0\delta(\phi-\phi_0)$, where $\phi_0$ 
is randomly selected from 5 to 500~rad~m$^{-2}$ and 
${\rm SNR}=|F_0|/\sigma\approx32$. This corresponds to the Faraday screen model 
with only one component in Sect.~\ref{sect:qu-fitting}. The fractional 
polarized intensity 
is constant over $\lambda^2$ and the polarization angle is linearly related to 
$\lambda^2$. An example (Model 1, Table~\ref{tbl:all}) is shown in 
Fig.~\ref{fig:fdf_thin}. Pulsars and some extragalactic radio sources are in 
this category.   

\item {\it Two Faraday thin components.} The Faraday spectrum is 
written as $F(\phi)=F_1\delta(\phi-\phi_1)+F_2\delta(\phi-\phi_2)$, where 
the ratio of the amplitudes is between 0.35 and 0.95 and the strong  
component has an SNR of 32, 
$0\degr\leq|\chi_1-\chi_2|<180\degr$, and 
15~rad~m$^{-2}$$\leq|\phi_1-\phi_2|\leq$140~rad~m$^{-2}$. The separations of 
the two components in Faraday depth for the various tests were both less than 
and greater than $\Delta\phi_{\rm FPSF}\approx 122$~rad~m$^{-2}$. 
\citet{ms96} measured RMs across 23 extended radio sources 
around Galactic latitude $|b|=20\degr$ and found an rms scatter of 
$\sim$4$\pm$0.6~rad~m$^{-2}$ on angular scales $\sim1.2\arcmin$. This agrees 
with the results from a sample of 208 double sources with separations less than 
$3\arcmin$ at $|b|>25\degr$ from Rudnick et al. (2014, in preparation), who 
found a median difference of $\sim4$~rad~m$^{-2}$ between lobes. We therefore 
included small values of $|\phi_1-\phi_2|$, such as 14~rad~m$^{-2}$ for Model 
11 in Table~\ref{tbl:all}, even though they are less than the width of the FPSF.
 In the current tests, the Faraday depth separation is about 14~rad~m$^{-2}$ for 
Model 11, 40~rad~m$^{-2}$ for Models 5, 6, 8, and 9, 80~rad~m$^{-2}$ for 
Models 7 and 13, 140~rad~m$^{-2}$ for Models 4 and 10, and 180~rad~m$^{-2}$ 
for Model 12 (Table~\ref{tbl:all}). We did not 
choose the models specifically to represent situations with RM ambiguities 
\citep[e.g.][]{kai+14}. 
The spectral 
indices ($I_\nu\propto\nu^{-\alpha}$) are either the same for the two 
components, or set to $\alpha_1=0$ and $\alpha_2=0.7$, with an assumption that 
each component has a constant fractional polarization as a function of 
$\lambda^2$. This model represents, e.g., an unresolved double source such as a radio 
galaxy with either two lobes or one core and one jet, each with its own Faraday 
screen. 
The polarized intensity and polarization angle can be calculated from 
Eq.~(\ref{eq:fs3}). An example (Model 4, Table~\ref{tbl:all}) is shown in 
Fig.~\ref{fig:fdf_thin}.   

\begin{figure}
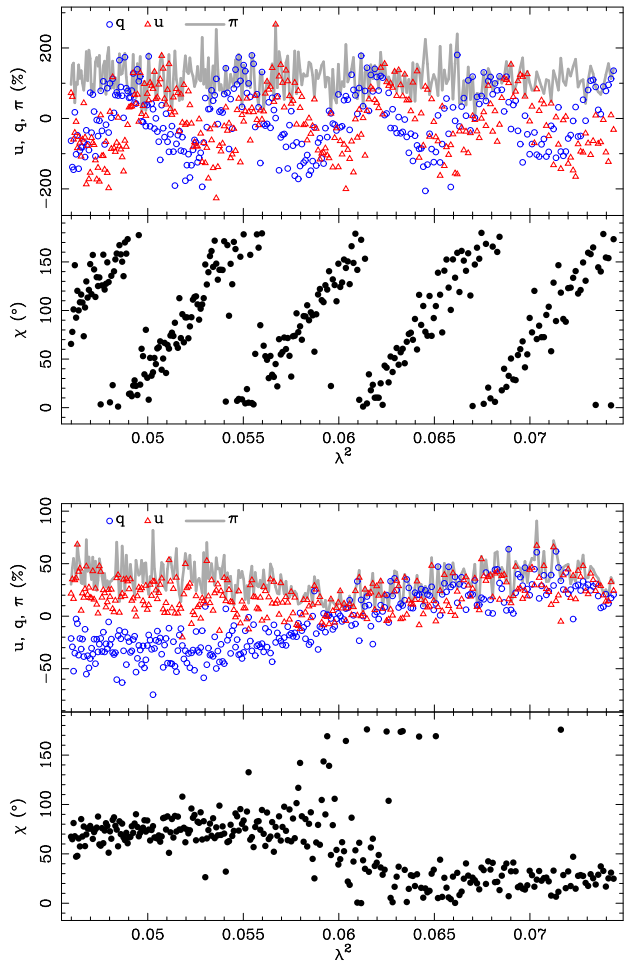

\centering
\includegraphics[angle=-90,scale=0.35]{f1a_c.ps}\\[4mm]
\includegraphics[angle=-90,scale=0.35]{f1b_c.ps}
\caption{Examples of benchmark test data with input $F(\phi)$ based on one 
Faraday thin component (upper, Model 1 in Table~\ref{tbl:all}) and based on two 
Faraday thin components with $\Delta\phi=141$~rad~m$^{-2}$ (lower, Model 4 in 
Table~\ref{tbl:all}). The red triangles display $u=U/I$, blue circles display $q=Q/I$,
 grey lines display $\pi=\sqrt{u^2+q^2}$ and the dots display polarization 
angles. The point-to-point scatter reflects the noise added to each point.}
\label{fig:fdf_thin}
\end{figure}

\item {\it Faraday thick component.} The Faraday spectrum is 
described by a boxcar function: $F(\phi)=F_0$ for 
$\phi_c-\phi_s/2<\phi<\phi_c+\phi_s/2$; 
$F(\phi)=0$ elsewhere. Here $\phi_c$ is the central Faraday depth of 
the component, $\phi_s$ is the Faraday depth extent of the component, and  
${\rm SNR}=|F_0|\times\phi_s/\sigma\approx32$. 
This corresponds 
to the slab model described in Sect.~\ref{sect:qu-fitting}, 
where $\phi_0$ is 
either $\phi_c+\phi_s/2$ or $\phi_c-\phi_s/2$. 
The values of $\phi_s$ were 25 and 50~rad~m$^{-2}$, 
both less than the maximum scale of 68~rad~m$^{-2}$, although for the latter 
case there will still be strong depolarization at the lower frequencies. The 
thick component approximately depicts a source with both a synchrotron-emitting 
and Faraday rotation medium, such as Galactic diffuse emission, supernova 
remnants and nearby galaxies. In reality, the distribution of polarized 
emission is probably more extended than that of thermal gas, and thus the 
Faraday spectra can deviate substantially from a boxcar shape 
\citep{bfss12}. An example (Model 14, Table~\ref{tbl:all}) is shown in 
Fig.~\ref{fig:fdf_thick}. In addition, three of the tests included both a 
Faraday thin and Faraday thick component.
\end{itemize}

\begin{figure}
\centering
\includegraphics[angle=-90,scale=0.35]{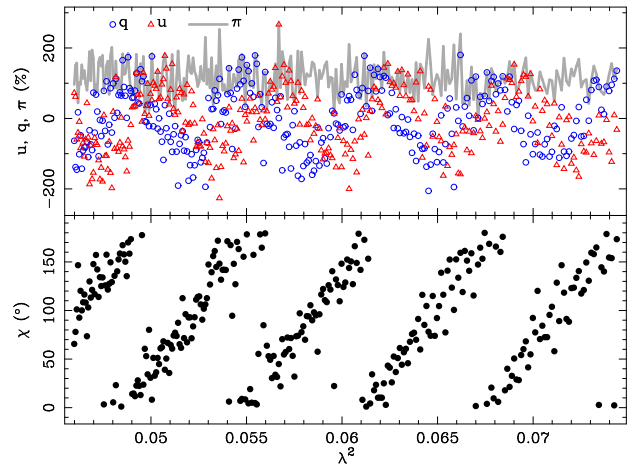}
\caption{The same as Fig.~\ref{fig:fdf_thin} but for the Faraday thick model 
with a Faraday depth extent $\phi_s=25$~rad~m$^{-2}$ (Model 16 in 
Table~\ref{tbl:all}).}
\label{fig:fdf_thick}
\end{figure}

\section{Benchmark results}\label{results}

\subsection{Overall performances of all the methods}
All the results from the data challenge are listed in Table~\ref{tbl:all}. 
We first define three quantities as figures of merit to assess these results. 
All of these quantities are driven by the science goals of POSSUM 
\citep{glt10}. 

The first figure of merit is how well the weighted average of Faraday depth, 
defined as
\begin{equation}
{\rm RM_{wtd}}=\frac{\sum_i |F_i|\phi_i}{\sum_i|F_i|},
\end{equation}
where $i$ denotes different components, 
meets that of the input model. This is motivated by the science goal of using 
an RM grid to study foreground magnetic fields. Ideally, sources with only one 
Faraday thin component are best suited for this purpose, in which case 
RM$_{\rm wtd}$=RM. Sources with more Faraday thin components or Faraday thick 
components can also be used at the risk of increasing the scatter between the 
RMs. The average is weighted by polarized intensity because in the limit that 
the two components are not well-resolved in Faraday space, it is similar to the 
average result that one would get from fitting a single component to the data. 
The best single component fit in any individual case also depends on the 
relative phase (namely the difference of polarization angles) of the two input components. 
The theoretical error ($1\sigma$ uncertainty) of the Faraday depth for 
a component from the Faraday synthesis method can be written as
$\sigma_{\rm RM}=0.5\Delta\phi_{\rm FPSF}/{\rm SNR}$, which was previously  
used by e.g. \citet{mgh+10} and has been verified by us with Monte-Carlo 
simulations. In the current tests, $\rm SNR=32$, the theoretical error is thus 
about 1.9~rad~m$^{-2}$. 
For $\rm RM_{wtd}$ the error depends on the relative amplitudes of the 
components and is between $\sigma_{\rm RM}/\sqrt{2}$ and $\sigma_{\rm RM}$. For 
the discussions below we simply use $\sigma_{\rm RM}$ for comparison with the 
results. The expected median of $\rm |RM_{wtd}\,(test-model)|$ and the error of 
the median are then about $0.675\sigma_{\rm RM}\approx1.3$~rad~m$^{-2}$ and 
$\sigma_{\rm RM}/\sqrt{N_{\rm test}}\approx0.5$~rad~m$^{-2}$, respectively. 
Here $N_{\rm test}=17$ is the number of tests.

The second figure of merit is the comparison between the fit and model values 
of the separation of the two Faraday thin components or the width of a Faraday 
thick component, calculated as
\begin{equation}
\Delta\phi=|\phi_1-\phi_2|.
\end{equation}
If there is only one Faraday thin component, we define $\Delta\phi=0$. 
For a source with two Faraday thin components, $\Delta\phi$ measures the 
Faraday depth difference between these two components, which is an indicator of the 
thermal environment local to or inside the source. The ability to measure 
$\Delta\phi$ is crucial for exploring magnetic fields in the cosmic web 
\citep{aktr14}. For a source with a Faraday thick component, such as  Galactic 
diffuse polarized emission, $\Delta\phi=\phi_s$ (Table~\ref{tbl:all}), 
can be modelled in terms of a rough 
spatial scale of the structure particularly towards high Galactic latitudes. 
The fit value for $\Delta\phi$ also indicates whether a method is able to 
recognize the complexity of $F(\phi)$ and avoid some of the RM ambiguities 
\citep{frb11,kai+14}. 

For the second figure of merit, we therefore list for 
each method the difference between the model inputs and test 
outputs for the separation or width $|\Delta\phi\,({\rm test-model})|$. 
The theoretical error can be estimated as 
$\sigma_{\Delta\phi}=\sqrt{2}\sigma_{\rm RM}\approx2.7$~rad~m$^{-2}$, 
assuming that the two components do not interfere with each other in 
the Faraday spectrum. 
The 
theoretical median of $|\Delta\phi\,({\rm test-model})|$ and the error of the 
median are about $0.675\sigma_{\Delta\phi}\approx1.8$~rad~m$^{-2}$ and 
$\sigma_{\Delta\phi}/\sqrt{N_{\rm test}}\approx0.7$~rad~m$^{-2}$, 
respectively. Since only the strong component has an SNR 32, the 
theoretical median of $|\Delta\phi\,({\rm test-model})|$ is slightly lower than the true value.

The last figure of merit is the reduced chi-squared $\chi_r^2$ calculated as
\begin{equation}
\chi_r^2=\frac{1}{2N-3N_c}\sum_{i=1}^{N}\frac{\left[(\widetilde{U}_i-U_i)^2+(\widetilde{Q}_i-Q_i)^2\right]}{\sigma_i^2},
\end{equation}
where $N$ is the number of frequency channels, $N_c$ is the number of 
components, $\sigma_i=1$ is the rms noise in each channel, and $\widetilde{U}$ 
and $\widetilde{Q}$ are model values. For $N=300$, the expected mean of 
$\chi_r^2$ is $\sim$1 with a scatter of $\sim$0.02. In addition to measuring 
the accuracy of the fitted Faraday depth(s), $\chi_r^2$ also checks whether 
polarized amplitudes and phases are properly solved, which would be important 
for detailed studies of individual sources.  

\begin{figure}
\centering
\includegraphics[angle=-90,scale=0.35]{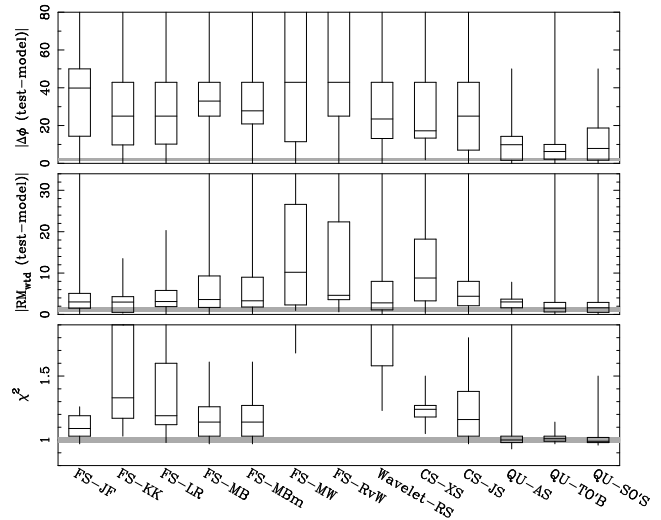}
\caption{Box-and-whisker plots for figures of merit over all 17 tests. 
The boxes show the first, 
the second (median) and the third quartile, and the ends of the whiskers show 
the minimum and maximum values. The shaded areas indicate $1\sigma$ range above 
and below the theoretical medians. See Table~\ref{tbl:summary} for the medians 
and uncertainties. }
\label{fig:med}
\end{figure}

The three figures of merit, $\chi_r^2$, $\rm |RM_{wtd}\,(test-model)|$ and 
$|\Delta\phi\,({\rm test-model})|$, for all of the methods are shown in 
Fig.~\ref{fig:med}. The median values for these figures of merit for each 
fitting procedure as well as the theoretical expectations are listed in 
Table~\ref{tbl:summary}. Although there are always outliers for each method, 
clear trends can be recognized for each figure of merit: (1) All the 
$QU$-fitting methods have $\chi_r^2\sim$1 and all the other methods have median 
values of $\chi_r^2$ significantly larger than the expectation. 
$QU$-fitting allows only 3 or 6 parameters (1 or 2 Faraday components), and then minimizes $\chi_r^2$. The larger values of $\chi_r^2$ for the other 
methods may be due to the fact that they allowed signal power to be spread over a much larger number of components, but were forced to report only the strongest 1 or 2 components, which, by themselves, would produce a poor $\chi_r^2$.

Note that FS-JF used the smallest sampling of 1~rad~m$^{-2}$ in the 
Faraday depth domain and obtained the best $\chi_r^2$ among the Faraday 
synthesis methods. Further tests are needed to investigate how the 
sampling influences $\chi_r^2$. 
(2) Only two of the $QU$-fitting methods, 
QU-TO'B and QU-SO'S have a median $\rm |RM_{wtd}\,(test-model)|$ consistent 
with the (conservative) theoretical values. (3) The $QU$-fitting methods have a 
much smaller median $|\Delta\phi\,({\rm test-model})|$ than other methods. 
However, none of the methods can reproduce the separations of the components 
within the idealized theoretical errors. 

\begin{table}[!htbp]
\scriptsize
\begin{center}
\caption{Median values for $\chi_r^2$, $\rm RM_{\rm wtd}$ and 
$\Delta\phi$ over all 17 tests. \label{tbl:summary}}
\begin{tabular}{cccc}
\hline\hline
Method & $\chi_r^2$ & $|$RM$_{\rm wtd}$ (test-model)$|$ & 
$|\Delta\phi$ (test-model)$|$\\
&&rad~m$^{-2}$&rad~m$^{-2}$\\\hline
\bf Expected & $\mathbf{1\pm0.02}$ & $\mathbf{1.3\pm0.5}$ & $\mathbf{1.8\pm0.7}$\\ 
FS-JF        &  1.09    &    3.0   &    39.9 \\
FS-KK        &  1.33    &    3.0   &    25.0 \\
FS-LR        &  1.19    &    3.1   &    25.0 \\
FS-MB        &  1.14    &    3.6   &    33.0 \\
FS-MBm       &  1.14    &    3.3   &    27.8 \\
FS-MW        &  3.95    &   10.2   &    42.9 \\
FS-RvW       & \nodata  &    4.6   &    42.9 \\
Wavelet-RS    &  2.01    &    2.8   &   23.5 \\
CS-XS         &  1.24    &    8.8   &   17.2 \\
CS-JS         &  1.16    &    4.4   &   25.0 \\
QU-AS         &  1.00    &    3.0   &   9.9 \\
QU-TO'B       &  1.01    &    1.5   &   6.3 \\
QU-SO'S       &  0.99    &    1.6   &   7.9 \\
\hline
\end{tabular}
\end{center}
\end{table}

Among the Faraday synthesis methods, FS-JF, FS-KK, FS-LR 
and FS-RvW all used similar algorithms but delivered different results. This is 
likely because these methods involved different processes of searching for 
peaks from the Faraday spectra and deciding which peaks to report, and these 
processes are very subjective. In the extreme case such as FS-RvW, all the 
components were picked out manually, as an automatic way was not ready during 
the tests.

The actual science from Faraday structures usually relies on only one 
parameter such as $\rm RM_{wtd}$ or $\Delta\phi$. However, more parameters need 
to be fit to achieve $\chi_r^2\sim1$. Therefore one cannot simply differentiate 
all the methods solely based on $\chi_r^2$. For most of the methods except for 
$QU$-fitting, although the $\chi_r^2$ values are much larger than 1 indicating 
very poor fits, the median values of $\rm |RM_{wtd}\,(test-model)|$ are within 
about 5~rad~m$^{-2}$. These values deviate from the expectation, but are much 
less than the intrinsic Faraday depth fluctuations over degree scales from the 
Galaxy~\citep{sr09,sch10}. This means $\rm RM_{wtd}$ values from these methods 
may still be suitable for studying magnetic fields in Galactic objects. 

\begin{figure}
\centering
\includegraphics[scale=0.25]{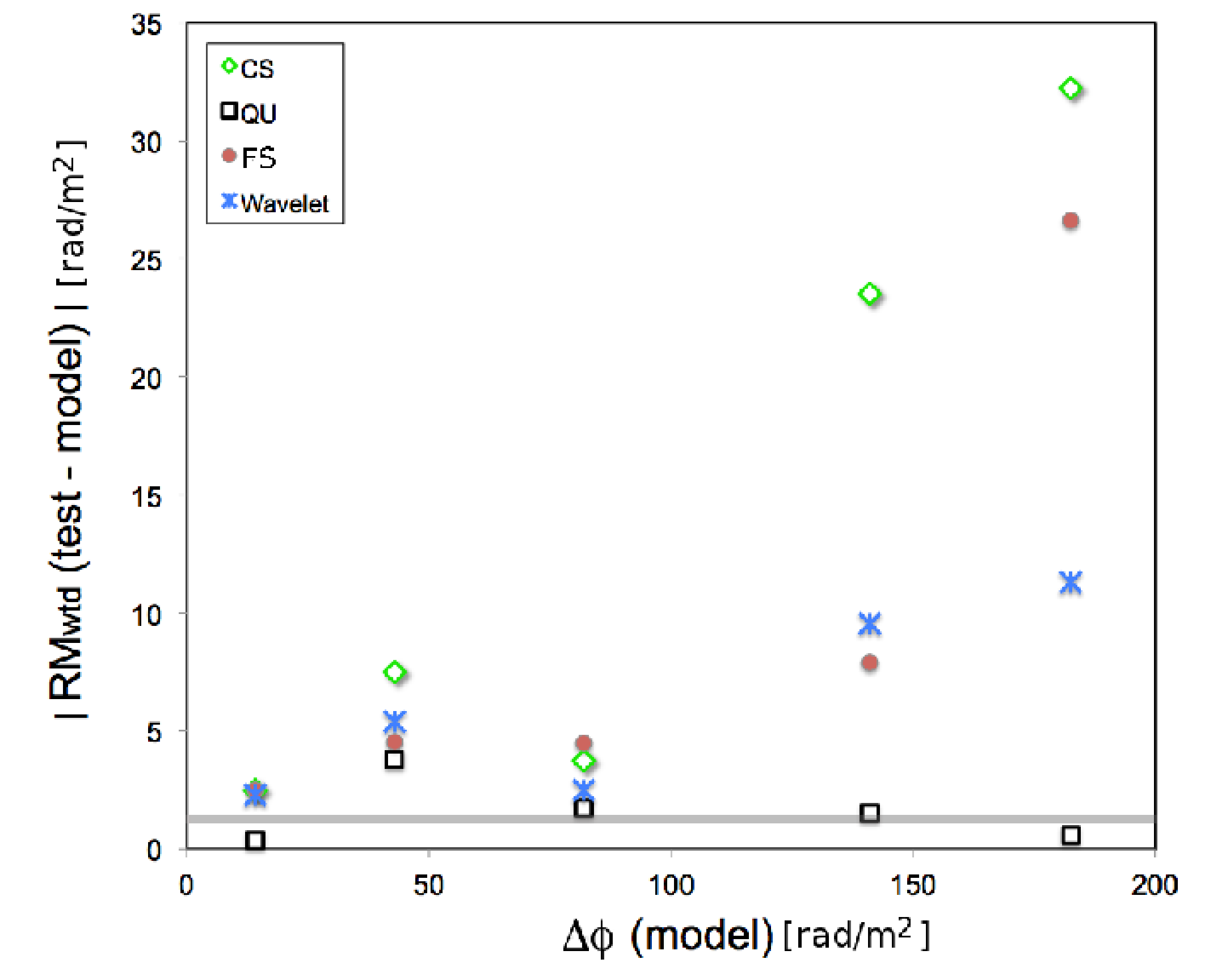}\\[8mm]
\includegraphics[scale=0.30]{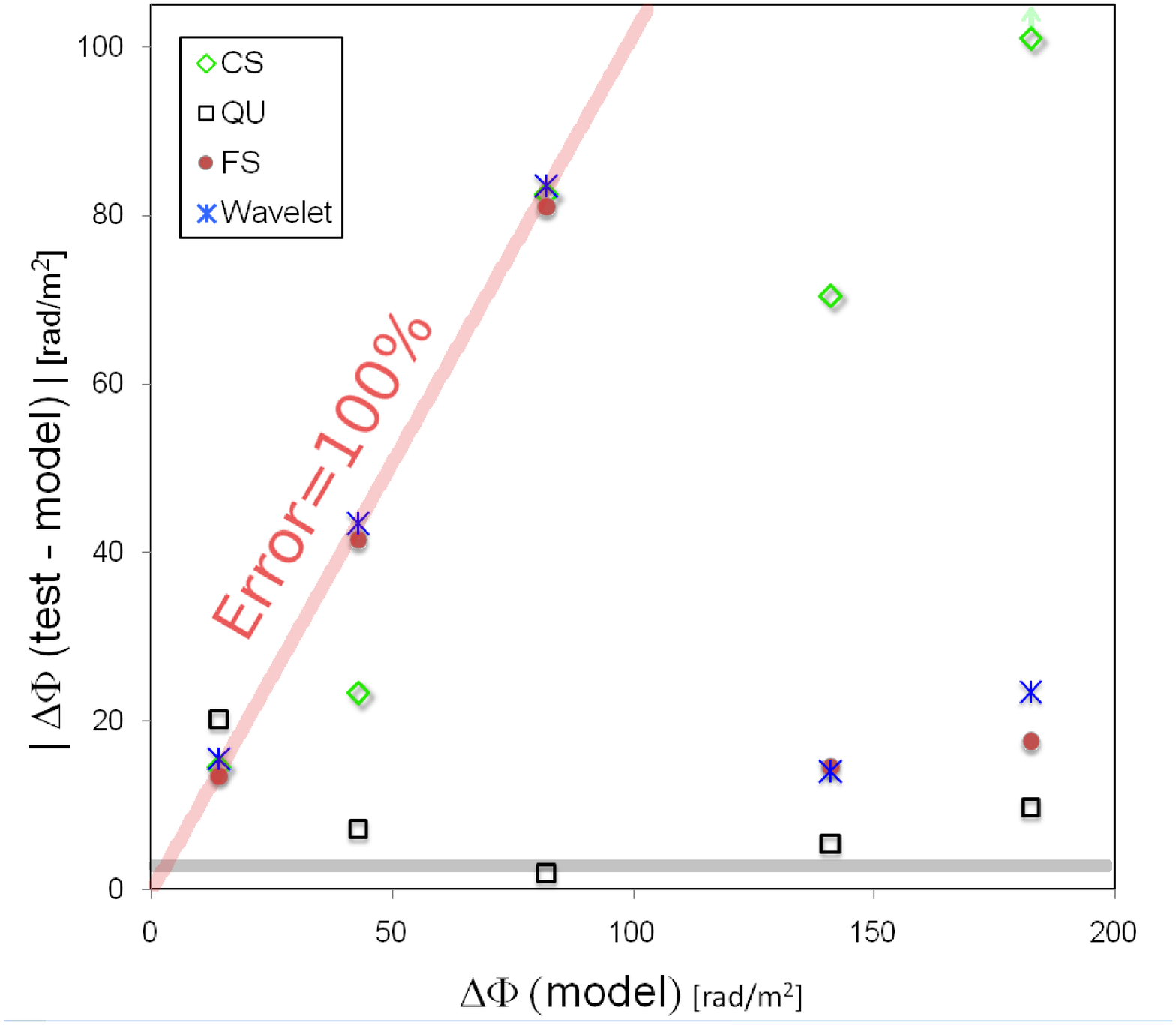}
\caption{$\rm |RM_{wtd}\,(test-model)|$ (upper panel) and 
$|\Delta\phi\,({\rm test-model})|$ (lower panel) versus model $\Delta\phi$ for 
two Faraday thin components, showing the median results averaging over all 
algorithms of each type (Faraday synthesis, wavelets, compressive sampling and 
$QU$-fitting) and averaging all the  tests with similar values of input 
$\Delta\phi$. The grey shaded regions at the bottom of each panel show the 
median expected values from the SNR alone (Table~\ref{tbl:summary}).}
\label{fig:thin_rmwtd_drm}
\end{figure}

\subsection{Fitting accuracies for different $F(\phi)$ models}

We summarize here the performance of the various methods for each type of input 
model. For a single Faraday thin component model, most of the methods reproduce 
the inputs well. However, FS-MW and CS-XS identify more than one component for 
one or two cases, and FS-MW and FS-RvW obtain a too high $\rm RM_{wtd}$ for one 
case. 

For two Faraday thin components, the results are much less satisfactory. 
Fig. ~\ref{fig:thin_rmwtd_drm} summarizes the results, now averaging over all 
algorithms of each type (Faraday synthesis, wavelets, compressive sampling, and 
$QU$-fitting). We plot the median deviations from the models as a function of 
the separation between the two components, $\Delta\phi$.  

Looking first at  $\rm |RM_{wtd} (test-model)|$, we see that {\em only} 
$QU$-fitting produces results that deviate from the input model approximately 
as expected. Somewhat surprisingly, the deviations for non-$QU$-fitting methods 
increase sharply when the separation between components is larger than 
$\Delta\phi_{\rm FPSF}$, reaching more than an order of magnitude worse than 
expected. The reason is unclear. More tests covering a full range of 
amplitude ratio and phase difference are needed to investigate this.
For $\Delta\phi$ values smaller than the $\Delta\phi_{\rm FPSF}$, 
$QU$-fitting produces results that are about a factor of two better than other 
algorithms.

\begin{figure}
\centering
\includegraphics[scale=0.28]{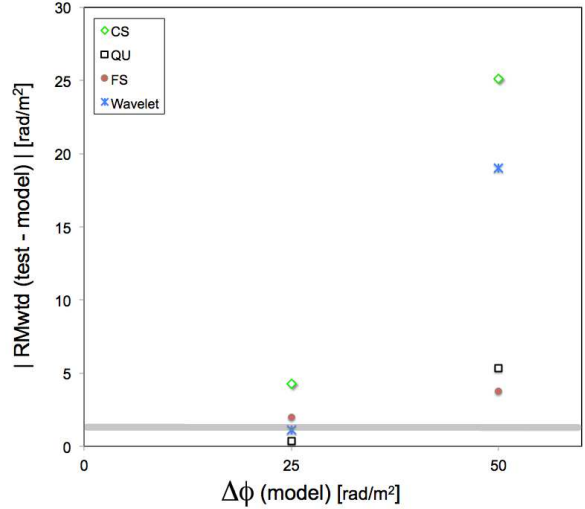}\\[8mm]
\includegraphics[scale=0.28]{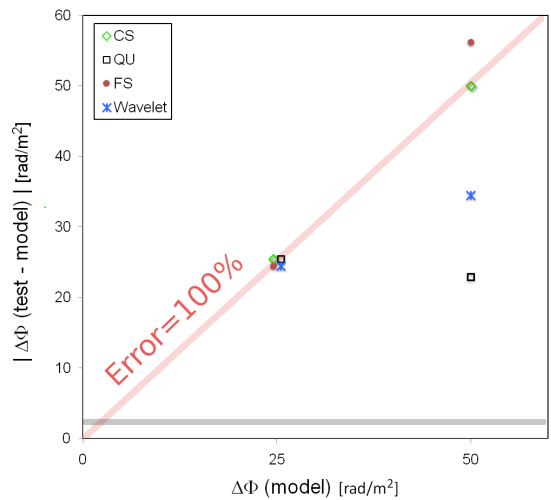}
\caption{Similar as Fig.~\ref{fig:thin_rmwtd_drm} but for the Faraday thick 
model.} 
\label{fig:thick_rmwtd_drm}
\end{figure}

Turning now to measurements of $\Delta\phi$ itself, $QU$-fitting results 
generally identify the existence of more than a single Faraday thin component, 
but deviate from the input model significantly more than expected 
theoretically. This is likely an irreducible error introduced by the multiple 
degrees of freedom in the fits, as discussed below. At values of $\Delta\phi$ 
much smaller than $\Delta\phi_{\rm FPSF}$, all the methods including 
$QU$-fitting have 100\% errors, and this is likely limited by the current 
bandwidth of 300~MHz. At larger values of $\Delta\phi$ but still less than 
$\Delta\phi_{\rm FPSF}$, the other methods still produce 100\% errors 
indicating that they are unable to identify the 
existence of two components. At even larger separations of 
$\Delta\phi_{\rm FPSF}<\Delta\phi\lesssim1.5\Delta\phi_{\rm FPSF}$, 
Faraday synthesis and wavelet algorithms at least recognize the two components, 
but do a poor job of measuring their separations compared to $QU$-fitting. 
Actually the Faraday synthesis method could fail to identify the two components 
if the intensity ratio of the two components is much smaller than that in the 
current tests \citep{kai+14}. 

For tests with Faraday thick components, the $\rm |RM_{wtd} (test-model)|$ 
and $|\Delta\phi\,({\rm test-model})|$ against the model $\Delta\phi$ are 
shown in Fig.~\ref{fig:thick_rmwtd_drm}. $QU$-fitting produces an 
$\rm RM_{wtd}$ consistent with the theoretical expectation for 
$\Delta\phi=25$~rad~m$^{-2}$. For $\Delta\phi=50$~rad~m$^{-2}$, Faraday 
synthesis is slightly better than $QU$-fitting but both produce $\rm RM_{wtd}$ 
deviating largely from the input. For the measurements of 
$\Delta\phi$, all the methods fail to correctly recognize the thick component 
when $\Delta\phi=25$~rad~m$^{-2}$. For $\Delta\phi=50$~rad~m$^{-2}$, all the 
methods are able to identify the thick component, although they cannot 
reproduce $\Delta\phi$ properly. 

\subsection{Extended tests with $QU$-fitting methods}

A series of more demanding two component tests was run with lower SNR and 
smaller $\Delta\phi$. These tests were run only with $QU$-fitting methods, 
because of their low $\chi_r^2$ in the experiments described above. The 
series consisted of eight tests, each with two components of equal polarized 
and equal total intensity amplitudes. These parameters were known to each 
participant prior to the fitting. The SNR is 15 for each component, which is still at 
least a factor of 1.5 higher than envisioned for, e.g., the POSSUM detection 
limits~\citep[e.g.][]{gsk12}. 

Two of the $QU$-fitting methods (QU-SO'S and QU-AS) used here were described as 
before. In the time between the two rounds of tests, three additional 
$QU$-fitting routines became available. First is 
QU-SI\footnote{SI=Shinsuke Ideguchi}, which uses a MCMC algorithm similar 
to QU-AS. The second new routine is RMFIT in AIPS, written by E. Greisen (NRAO),
 which is an interactive Levenberg-Marquardt, non-linear least squared 
minimization program working pixel by pixel on $QU$ frequency cubes. The number 
of components to be fit is specified beforehand (two in this case) and optional 
starting guesses can be made based on the Faraday spectrum. This method is 
designated QU-LR as the tests were run by L. Rudnick. Finally, the third new 
routine is QU-JB\footnote{JB=Justin Bray}, which uses 
MultiNest~\citep{fh08, fhb09}, a library implementing a nested sampling 
algorithm. This algorithm iteratively selects more closely-clustered points 
around identified peaks in the likelihood/posterior distribution, which makes 
it more robust against multimodal distributions or curving degeneracies than 
traditional MCMC methods (although the test cases here generally did not appear 
to possess these properties). It calculates the local evidence for Bayesian 
model selection purposes, but for these extended tests the models were known, 
so it was used simply to calculate the maximum-likelihood fit, and should 
produce similar results to any other maximum-likelihood $QU$-fitting method.

\begin{table}
\scriptsize
\begin{center}
\caption {Results from $QU$-fitting methods for the extended tests with SNR 15 
and model $\Delta\phi=20,\,40$~rad~m$^{-2}$. Summary performance of the various 
methods according to the adopted figures of merit: $\rm RM_{\rm wtd}$ and 
$\Delta\phi$. \label{tbl:qu_summary}}
\begin{tabular}{c|rrrrrr}
\tableline\tableline
&Model & QU-LR & QU-SI & QU-SO'S & QU-AS & QU-JB\\
\tableline
& $-$140  &  $-$138   &  $-$146   & $-$138  &  $-$138  &  $-$138\\ 
& $-$140  &  $-$136   &  $-$145   & $-$136  &  $-$136  &  $-$135\\ 
& $-$160  &  $-$165   &  $-$156   & $-$164  &  $-$165  &  $-$166\\ 
$\rm RM_{wtd}$& $-$160  &  $-$163   &  $-$152   & $-$151  &  $-$162  &  $-$162\\ 
rad~m$^{-2}$& $-$10  &  $-$  7   &  19   & $-$10  &  $-$6  &  $-$6\\ 
&  180  &    155   &  182   & 187  &  188  &  186\\ 
&  180  &    181   &  181   & 157  &  181  &  182\\ 
&  110  &    110   &  112   & 110  &  111  &  111\\ 
\tableline
& 20    &     80       &  85     &   80     &   80  &  81 \\
& 20    &     23       &  77     &   28     &   23  &  30 \\
& 20    &     24       &  33     &   19     &   19  &  26 \\
$\Delta\phi$& 20    &     62       &   6     &   20     &   61  &  61 \\
rad~m$^{-2}$& 40    &     43       & 286     &   25     &   43  &  43 \\
& 40    &    216       &  44     &   24     &   25  &  25 \\
& 40    &     46       &  36     &   54     &   45  &  42 \\
& 40    &     41       &  22     &   41     &   30  &  45 \\
\tableline
\end{tabular}
\end{center}
\end{table}

The results are shown in Table~\ref{tbl:qu_summary}. The $\chi_r^2$ values for 
all the tests are very close to 1 and are not listed in the table. The weighted 
RM averages 
are close to the input values, with a median difference of 4~rad~m$^{-2}$, 
within the expected errors. On the other hand, there are some very large 
differences between the fits and the models, {\em but with no obvious signature 
in the $\chi_r^2$ values to show if the fit is inappropriate.} The 
difference in the separation between the two fit components are also typically 
in good agreement with the model, but again, with occasional large 
discrepancies and no indication of high $\chi_r^2$. These discrepancies have an important effect when considering 
the reliability of derived Faraday parameters, and are discussed in more detail 
in Sect.~\ref{discussions}.

\section{Discussion}\label{discussions}

The studies of magnetic fields in foreground Galactic and intergalactic plasmas 
rely on properly determined values for $\rm RM_{wtd}$. The scatter in 
$\rm RM_{wtd}$ between independent extragalactic sources is a source of noise for such 
foreground experiments, and only by averaging over large numbers of sources can 
this number be reduced \citep[see][]{ro14}. This, in turn, limits the smallest angular area for 
which sufficiently reliable average RMs can be determined. It is therefore 
critical that the errors for determining $\rm RM_{wtd}$ are much less than the 
intrinsic scatter between sources.

In a sample of 37 extragalactic sources, \cite{lgb+11} used Faraday synthesis 
and found that $\sim$25\% of them had more than a single Faraday component. 
Given our results that Faraday synthesis often misses composite structure, this 
must be considered a very conservative lower limit. 
\citet{ghba13} also showed that a large fraction of the sources 
observed towards M~31 has composite Faraday spectra. 
\citet{fgc14} has recently 
compiled a catalog containing about 1000 extragalactic sources with 
fraction polarization measured at more than 2 independent frequencies, and a 
large majority of these sources shows depolarization behavior indicating they 
are composite. Wherever there is Faraday complexity, only $QU$-fitting methods 
will produce results that do not introduce extraneous scatter in 
$\rm RM_{wtd}$, given the algorithms available today. 
It is important to note 
that these conclusions all apply where $\lambda_{\rm max}^2$ is only 1.6 times 
$\lambda_{\rm min}^2$ and where $\delta\phi$ is often comparable to or less 
than $\Delta\phi_{\rm FPSF}$. These are realistic conditions, and important to 
understand, but results from tests with much broader bandwidths and/or broader 
Faraday structures would give different results. 

For one of the original suite of tests (Model 13, Table~\ref{tbl:all}), and one 
of the extended, more demanding tests, we performed an additional 
experiment. We took the same model, and created 500 realizations of the noise, 
with the same rms, and then fit each of these using QU-JB.

\begin{figure}
\centering
\includegraphics[scale=0.28]{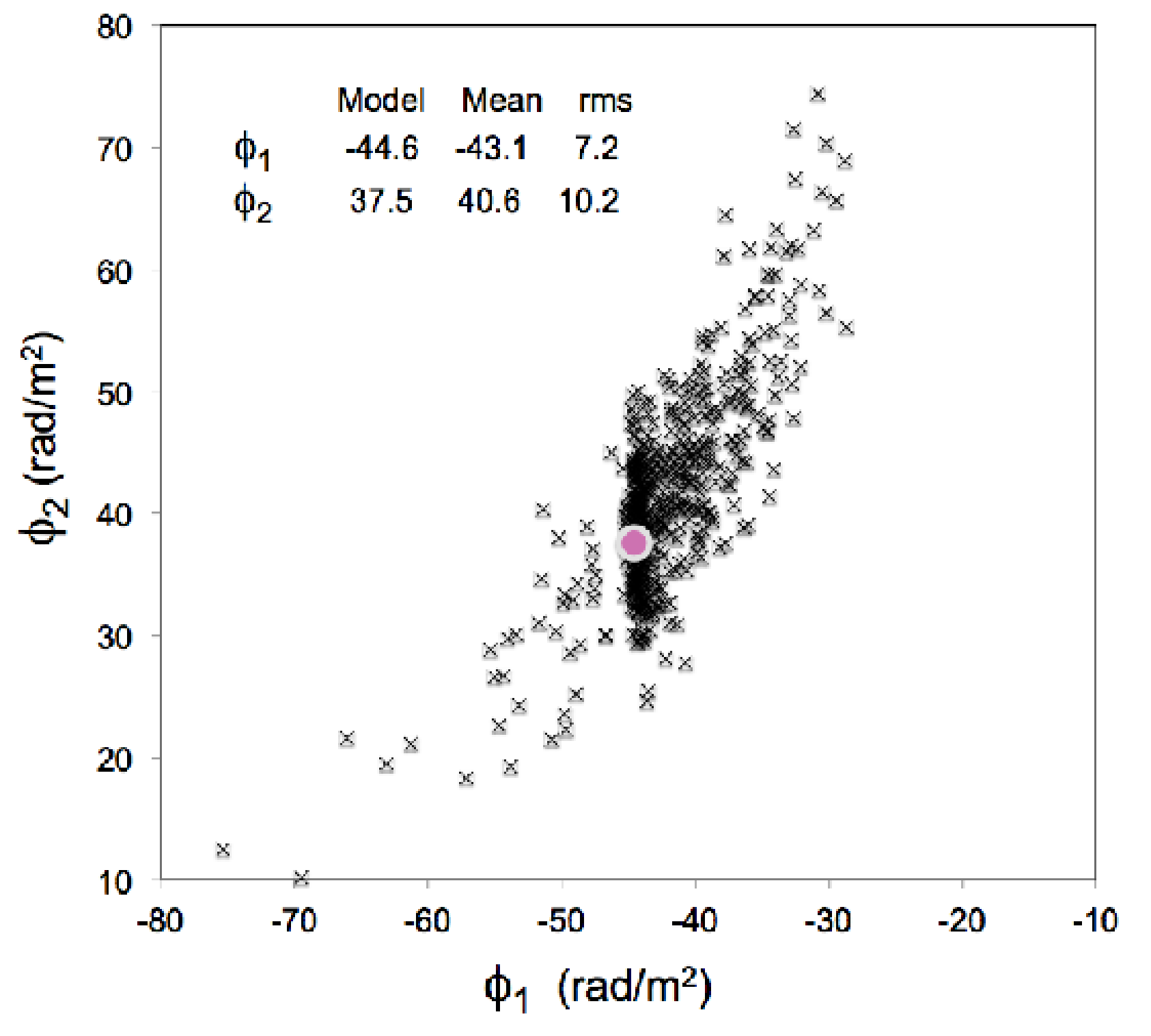}\\[4mm]
\includegraphics[scale=0.28]{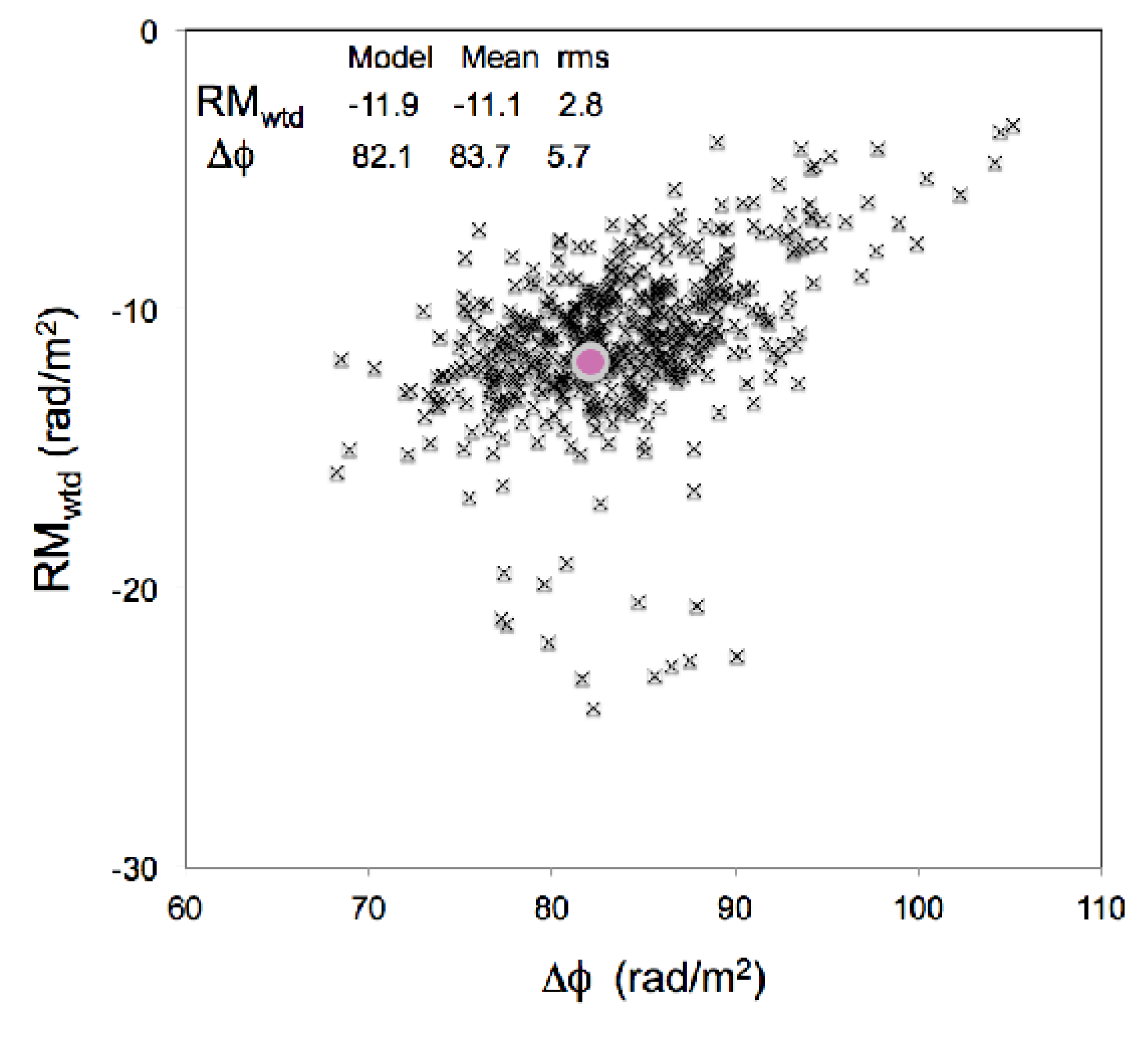}
\caption{$QU$-fitting results from 500 realizations of a two-component model 
with $\Delta\phi=82$~rad~m$^{-2}$ and an SNR of about 32. The filled circle 
indicates the model values. The input values and the mean and $1\sigma$ 
uncertainty of the outputs are also indicated.} 
\label{fig:errorcheck_q}
\end{figure}

Figure~\ref{fig:errorcheck_q} shows the distribution of the values of $\phi_1$ 
and $\phi_2$, along with the model for one test with an input 
$\Delta\phi\approx$82~rad~m$^{-2}$, somewhat smaller than 
$\Delta\phi_{\rm FPSF}$. The scatter in these values is 
7.2~rad~m$^{-2}$ (10.2~rad~m$^{-2}$), for the stronger (weaker) component, 
where the expected scatter for these components is only 
2~rad~m$^{-2}$ (3~rad~m$^{-2}$). The mean values also differ significantly from 
the input model. However, when we look at the primary measures of scientific 
interest, $\rm RM_{wtd}$ and $\Delta\phi$ (Fig.~\ref{fig:errorcheck_q}), we 
find these are much closer to the input models, with rms scatters of 2.8 and 
5.7~rad~m$^{-2}$ but still larger than the theoretical values of 1.7 and 
4.3~rad~m$^{-2}$.

\begin{figure}
\centering
\includegraphics[scale=0.28]{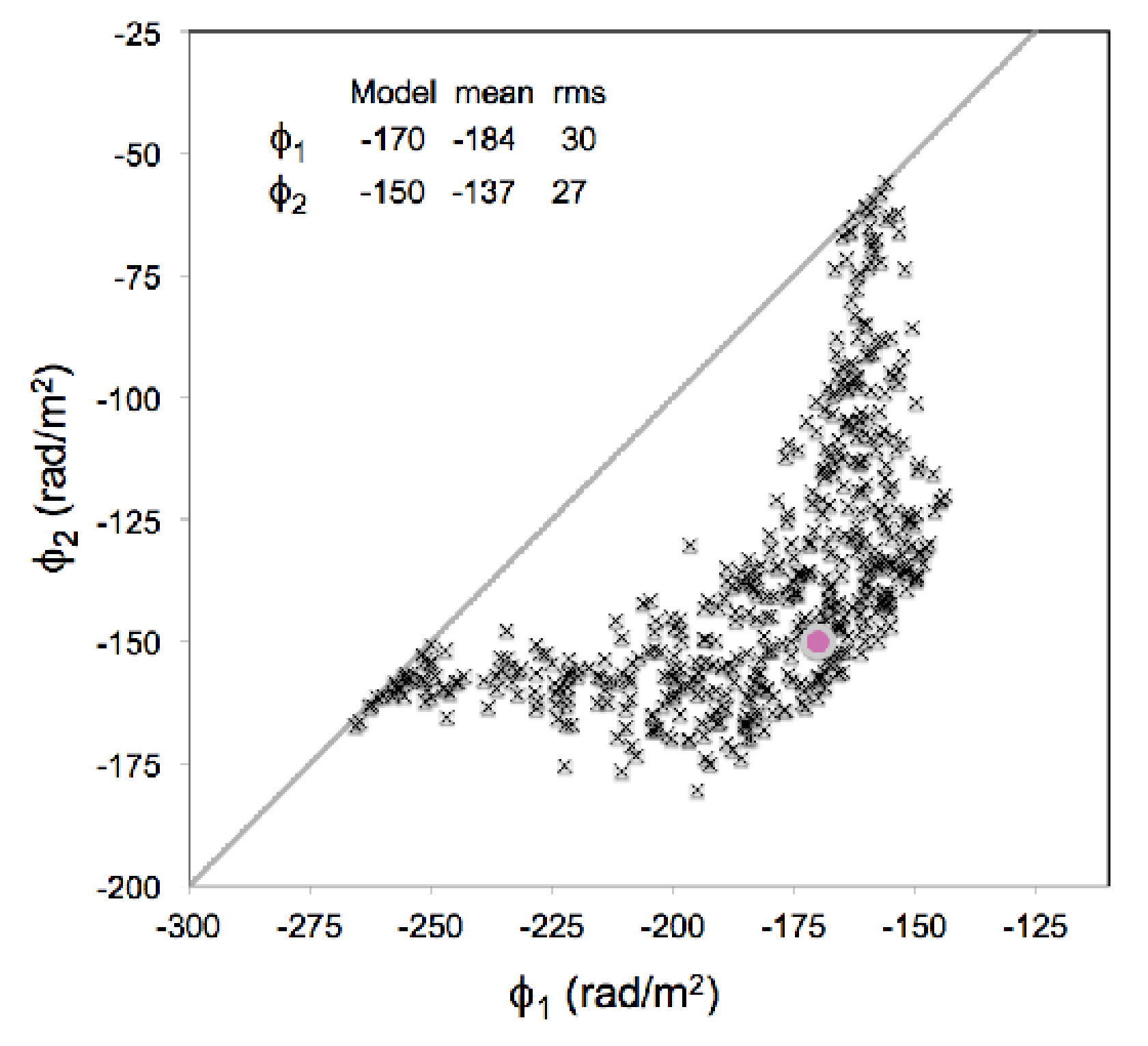}\\[4mm]
\includegraphics[scale=0.28]{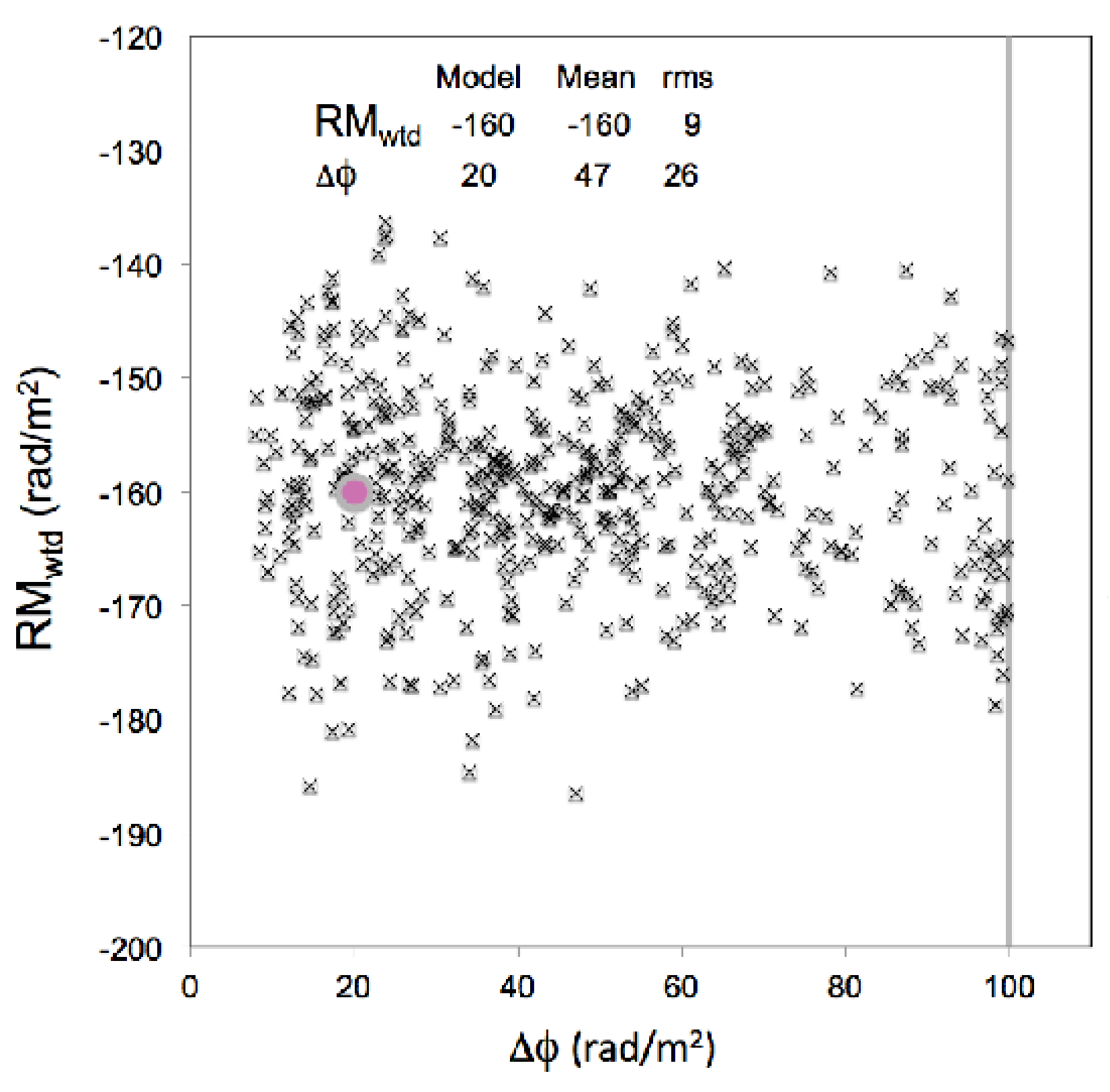}
\caption{The same as Fig.~\ref{fig:errorcheck_q} but for the extended 
test with a component separation of $\Delta\phi=20$~rad~m$^{-2}$ and an SNR of 
15.} 
\label{fig:errorcheck_xd}
\end{figure}

Performing the same experiment for a second test with an input $\Delta\phi$ of
$\sim$20~rad~m$^{-2}$ (Fig.~\ref{fig:errorcheck_xd}), we again see a very large 
scatter in $\phi_1$ and $\phi_2$, 30 and 27~rad~m$^{-2}$, where only 
4~rad~m$^{-2}$ was expected. We artificially constrained the solutions so that 
$\Delta\phi<100$~rad~m$^{-2}$, indicated by the grey line, otherwise the actual 
scatter would be even larger. Looking now at the $\rm RM_{wtd}$ 
(Fig.~\ref{fig:errorcheck_xd}), the solution is again more reasonable, with a 
scatter of only 9~rad~m$^{-2}$, where 3~rad~m$^{-2}$ was expected. Note that in 
none of these cases, however, do we approach the theoretical accuracy.

\begin{figure}
\centering
\includegraphics[scale=0.2]{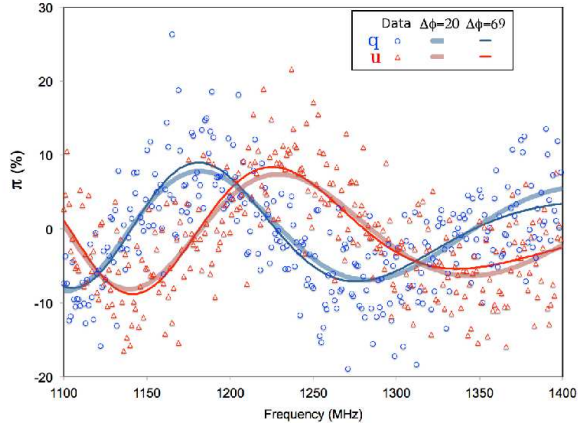}
\caption{Data, input model and an alternative model for one of the extended 
tests. $q$ (open circles) and $u$ (triangles) are plotted in units of 
percentage polarization as a function of frequency. The input model 
(thick lines) has $\Delta\phi$=20~rad~m$^{-2}$, while the alternative model 
(thin lines) has $\Delta\phi$=69~rad~m$^{-2}$, yielding approximately the 
same behaviors of $q$ and $u$.} 
\label{fig:uq_f_models}
\end{figure}

The situation is even worse when we look at $\Delta\phi$. Now we find 
a very broad range in possible values, and the mean is completely different 
from the input model. Again, we note the artificial cutoff at 
$\Delta\phi<100$~rad~m$^{-2}$. The reason for this very large scatter in 
$\Delta\phi$ can be seen in the plot of $q$ and $u$ versus $\lambda^2$ for the 
data and for two different model fits, shown in Fig.~\ref{fig:uq_f_models}. The 
fact that the $q$ and $u$ curves are almost identical means that these two very 
different fits cannot be distinguished, virtually independent of SNR, and 
independent of the fitting method. The degeneracy is reflected in the data 
themselves.

The above case presents a fundamental challenge to the accuracy of Faraday 
measurements. A more thorough investigation with different input phases and 
amplitudes for the two components are needed to understand the errors for both 
$\rm RM_{wtd}$ and $\Delta\phi$, and how often extreme cases such as the one 
above are expected to occur. 

\section{Conclusions}\label{conclusions}

We ran a data challenge to benchmark the current available algorithms to 
reconstruct Faraday spectra. The participating methods included 
Faraday synthesis and clean, wavelets, compressive sampling, and $QU$-fitting. 
The tests were carried out using a 1.1-1.4~GHz band, and signal-to-noise 
ratios of 32 and 15. Single and double Faraday thin components and Faraday 
thick components were tested. The figures of merit for these tests consisted of
the average Faraday depth weighted by polarized intensity $\rm RM_{wtd}$, the 
separation of two Faraday thin components or the extent of a Faraday thick 
component $\Delta\phi$, and $\chi_r^2$.

Our main results are:
\begin{itemize} 
\item Most methods are successful when only one Faraday thin component is 
present, with occasional failures where two components are found.
\item For composite Faraday spectra, the errors in $\rm RM_{wtd}$ are 
approximately as expected for $QU$-fitting, and much higher, sometimes by more 
than an order of magnitude, for all other methods.
\item Wherever $\Delta\phi>\Delta\phi_{\rm FPSF}$, the errors are much larger 
than expected, with $QU$-fitting performing the best. At values of $\Delta\phi$ 
smaller than $\Delta\phi_{\rm FPSF}$, only $QU$-fitting reliably recognizes the 
existence of two components. 
\item No methods, as currently implemented, work well for Faraday thick 
components mainly due to the low resolution in the Faraday depth domain 
given the current narrow bandwidth.
\item For certain combinations of input parameters, a wide variety of models 
are consistent with the same $QU$ data, and therefore no method will be 
able to accurately determine the true values.
\end{itemize} 

The problems identified above provide a lower limit to the uncertainties for 
all RM measurements currently in the literature. In fact, the previously published results 
will often be considerably worse than reported here because the RM 
determination method, i.e., fitting $\chi(\lambda^2)$ have additional 
uncertainties \citep{frb11}. Any results comparing the observed scatter in RMs, 
for example, to the theoretical uncertainties, will overestimate the 
contributions to the scatter from the sources themselves. 

These results have wide-ranging implications for future polarization surveys 
and their scientific applications. At present, the use of $QU$-fitting methods 
allows us to minimize the scatter in derived Faraday parameters and 
to reliably identify the presence of Faraday complexity. In addition, any 
attempts to separate observational errors from intrinsic RM variations between 
sources or from small-scale foreground fluctuations require a much more 
sophisticated exploration of error distributions than is currently available.  

Motivated by the problems found in recovering the Faraday structures, the 
participants are now trying to improve their methods. We are also developing 
more sophisticated models of the input $F(\phi)$ by including other 
depolarization mechanisms \citep[e.g.][]{fgc14} to closely represent the true source structures. The ASKAP early science program plans to cover the frequency 
range from 700 to 1800 MHz, which is much wider than the 300~MHz bandwidth for 
the data challenge reported here. All the methods will likely perform better in 
this new case. The GALFACTS survey at the frequency band of 1225 to 1525~MHz is 
producing polarization images of the Galactic diffuse emission which is 
Faraday thick. Clearly it is necessary to investigate further how all the 
methods are improved to reconstruct Faraday thick components. A second data 
challenge accounting for all the above effects is underway.

\begin{acknowledgements}
We thank Dr. Rainer Beck for suggesting the nomenclatures and Prof. Bryan 
Gaensler for very helpful comments on the manuscript. 
XHS was supported by the Australian Research Council through grant FL100100114.
 LR acknowledges support from the U.S. National Science Foundation through 
grant AST-1211595 to the University of Minnesota. JSF acknowledges the support 
of the ARC through grant DP0986386. TA and KK acknowledge the supports 
of the Japan Society for the Promotion of Science (JSPS). TO'B acknowledges 
support from the U.S. National Science Foundation Research Experience for 
Undergraduates grant to the School of Physics and Astronomy at the University 
of Minnesota. AS and JB gratefully acknowledge support from the European Research Council under grant ERC-2012-StG-307215 LODESTONE. SPO'S acknowledges the support of the Australian Research Council 
through grant FS100100033.
RS acknowledges support from the grant YD-520.2013.2 and 
benefited from the International Research Group Program conducted by the Perm 
region government. RJW is supported by NASA through the Einstein Postdoctoral 
grant number PF2-130104 awarded by the Chandra X-ray Center, which is operated 
by the Smithonian Astrophysical Observatory for NASA under contract NAS8-03060.
We acknowledge the assistance of Terry Thibeault, of the Minnesota Institute 
for Astrophysics, in blinding the tests for the participants. 
\end{acknowledgements}

\bibliography{/Users/xhsun/bibtex}

\clearpage
\LongTables

\begin{deluxetable*}{rrrrrrrrrrrr}
\tabletypesize{\scriptsize}
\tablecaption{Input models and results for the data challenge.\label{tbl:all}}
\tablehead{
&\multicolumn{3}{c}{Thin component 1}&\multicolumn{3}{c}{Thin component 2} &\multicolumn{3}{c}{Thick component}& & \\
& \colhead{$p_1$} & \colhead{$\phi_1$} & \colhead{$\chi_1$} & \colhead{$p_2$} & \colhead{$\phi_2$} &  \colhead{$\chi_2$}  &  \colhead{$p_0$} &
\colhead{$\phi_c$} & \colhead{$\phi_s$} &$\rm RM_{wtd}$ &$\chi_r^2$\\
& \colhead{\%} & \colhead{rad~m$^{-2}$} & \colhead{$\degr$} & \colhead{\%} & \colhead{rad~m$^{-2}$} & \colhead{$\degr$} & \colhead{\%} &
\colhead{rad~m$^{-2}$} & \colhead{rad~m$^{-2}$} & \colhead{rad~m$^{-2}$} &}
\startdata
\multicolumn{12}{c}{\normalsize \em One Faraday thin component}\\
\multicolumn{12}{c}{\hfill}\\
\bf Model 1  & \bf 100.00  & \bf 500.10  & \bf  40.00   &           &            &           &        &             &              & \bf   500.1  &\bf 0.98 \\
FS-JF      &  95.77  & 501.58  &   37.01   &           &            &           &        &             &              &    501.6  & 0.99 \\
FS-KK      &  27.30  & 500.60  &   40.50   &           &            &           &        &             &              &    500.6  & 1.76 \\
FS-LR      &  90.00  & 502.00  &   70.00   &           &            &           &        &             &              &    502.0  & 3.04 \\
FS-MB      &  95.60  & 501.5   &   37.3    &           &            &           &        &             &              &    501.5  & 0.99 \\
FS-MBn     &  95.60  & 501.6   &   37.4    &           &            &           &        &             &              &    501.6  & 0.99 \\
FS-MW      &  90.40  & 501.00  &   82.00   &    7.90   & $-$387.50  &  12.00    &        &             &              &    429.6  & 3.95 \\
FS-RvW     &         &         &           &           &            &           &        &   400.00    &              &    400.0  &  .   \\
Wavelet-RS &  13.00  & 499.00  &   42.00   &           &            &           &        &             &              &    499.0  & 2.12 \\
CS-JS      &  88.20  & 502.20  &   35.00   &           &            &           &        &             &              &    502.2  & 1.00 \\
CS-XS      &  61.54  & 496.12  &   45.00   &   10.80   &  498.02    &  45.00    &        &             &              &    496.4  & 1.25 \\
QU-AS      &  95.60  & 500.81  &   37.18   &           &            &           &        &             &              &    500.8  & 0.98 \\
QU-SO'S    &  96.00  & 501.4   &   37.5    &           &            &           &        &             &              &    501.4  & 0.99 \\
QU-TO'B    &  95.82  & 501.52  &   37.22   &           &            &           &        &             &              &    501.5  & 0.99 \\
\multicolumn{12}{c}{\hfill}\\
\bf Model 2     &\bf 100.00  &\bf  49.38  &\bf   60.00   &           &            &           &        &             &              & \bf    49.4  & \bf1.02 \\ 
FS-JF      &  96.20  &  52.35  &   50.20   &           &            &           &        &             &              &     52.4  & 1.02 \\
FS-KK      &  27.60  &  52.70  &   49.00   &           &            &           &        &             &              &     52.7  & 1.81 \\
FS-LR      &  91.00  &  52.50  &   50.00   &           &            &           &        &             &              &     52.5  & 1.02 \\
FS-MB      &  96.15  &  52.2   &   50.8    &           &            &           &        &             &              &     52.2  & 1.02 \\
FS-MBn     &  96.70  &  52.7   &   49.0    &           &            &           &        &             &              &     52.7  & 1.02 \\
FS-MW      &  92.30  &  50.50  & $-$4.90   &           &            &           &        &             &              &     50.5  & 5.56 \\
FS-RvW     &         &  50.00  &           &           &            &           &        &             &              &     50.0  &      \\ 
Wavelet-RS &  13.00  &  52.00  &   50.00   &           &            &           &        &             &              &     52.0  & 2.18 \\
CS-JS      &  89.60  &  52.60  &   48.80   &           &            &           &        &             &              &     52.6  & 1.03 \\
CS-XS      &  42.83  &  51.52  &   45.00   &   29.50   &   38.16    & $-$90.00  &        &             &              &     46.1  & 1.22 \\
QU-AS      &  96.15  &  52.41  &   49.85   &           &            &           &        &             &              &     52.4  & 1.02 \\
QU-SO'S    &  96.00  &  52.3   &   50.3    &           &            &           &        &             &              &     52.3  & 1.02 \\
QU-TO'B    &  96.44  &  52.35  &   50.28   &           &            &           &        &             &              &     52.3  & 1.02 \\
\multicolumn{12}{c}{\hfill}\\
\bf Model 3     &\bf 100.00  & \bf  4.96  & \bf  60.00   &           &            &           &        &             &              &   \bf   5.0  &\bf 0.97 \\ 
FS-JF      & 104.53  &   5.35  &   59.27   &           &            &           &        &             &              &      5.4  & 0.97 \\
FS-KK      &  29.50  &   5.80  &   58.00   &           &            &           &        &             &              &      5.8  & 1.90 \\
FS-LR      &  95.00  &   5.00  &  120.00   &           &            &           &        &             &              &      5.0  & 5.90 \\
FS-MB      & 106.04  &   4.6   &   61.6    &           &            &           &        &             &              &      4.6  & 0.97 \\
FS-MBn     & 104.40  &   5.0   &   60.3    &           &            &           &        &             &              &      5.0  & 0.97 \\
FS-MW      & 102.00  &   4.00  &$-$33.60   &           &            &           &        &             &              &      4.0  & 7.95 \\
FS-RvW     &         &   0.00  &           &           &            &           &        &             &              &      0.0  &      \\     
Wavelet-RS &  15.00  &   6.00  &   57.00   &           &            &           &        &             &              &      6.0  & 2.30 \\
CS-JS      &         &         &           &           &            &           & 124.00 &     4.57    &     18.30    &      4.6  & 0.98 \\
CS-XS      &  72.85  &   7.63  &   45.00   &   15.99   &  $-$5.72   &  90.00    &        &             &              &      5.2  & 1.09 \\
QU-AS      & 104.95  &   5.30  & $-$120.53 &           &            &           &        &             &              &      5.3  & 0.97 \\
QU-SO'S    &  11.50  &   6.3   &   56.3    &           &            &           &        &             &              &      6.3  & 0.98 \\
QU-TO'B    & 104.63  &   5.40  &   59.12   &           &            &           &        &             &              &      5.4  & 0.97 \\
\tableline
\multicolumn{12}{c}{\hfill}\\
\multicolumn{12}{c}{\normalsize \em Two Faraday thin components}\\
\multicolumn{12}{c}{\hfill}\\
\tableline
\bf Model 4     &\bf  25.00  &\bf $-$37.84 &\bf   0.00   & \bf  16.7    &\bf 103.18   &\bf $-$36.00    &        &             &              &   \bf  18.6  &\bf 0.97 \\
FS-JF      &  28.83  & $-$43.01 &  16.17   &   21.68   & 112.36   & $-$67.31   &        &             &              &     23.7  & 1.12 \\
FS-KK      &  32.90  & $-$43.70 &  19.50   &   17.20   & 111.40   & $-$64.50    &        &             &              &      9.5  & 1.14 \\
FS-LR      &  24.00  & $-$41.50 &  14.00   &   13.00   & 111.10   & $-$63.80    &        &             &              &     12.1  & 1.00 \\
FS-MB      &  24.31  & $-$47.4  &  31.5    &    1.09   & 121.4    & $-$99.0     &        &             &              &  $-$40.2  & 1.58 \\
FS-MBn     &  24.86  & $-$47.4  &  31.7    &    1.11   & 121.4    & $-$99.3     &        &             &              &  $-$40.1  & 1.56 \\
FS-MW      &  23.10  & $-$49.10 &  $-$3.00 &   15.20   &  95.70   & $-$90.00    &        &             &              &      8.4  & 6.92 \\
FS-RvW     &         & $-$50.00 &          &           &$-$100.00 &   0.00      &        &             &              &           &      \\
Wavelet-RS &  14.00  & $-$44.00 &  19.00   &    8.00   & 112.00   & $-$69.00    &        &             &              &     12.7  & 1.58 \\
CS-JS      &  21.50  & $-$42.80 &  15.10   &   12.60   & 111.40   & $-$64.40    &        &             &              &     14.2  & 1.04 \\
CS-XS      &  14.57  & $-$40.07 &   0.00   &    7.16   & $-$53.43 &    45.00    &        &             &              &  $-$44.5  & 1.19 \\
QU-AS      &  25.3   & $-$34.64 &  12.67   &   16.07   &  98.95   & $-$22.82    &        &             &              &     17.3  & 2.16 \\
QU-SO'S    &  25.00  & $-$40.7  &   8.6    &   16.00   & 108.2    & $-$54.1     &        &             &              &     17.4  & 0.96 \\
QU-TO'B    &  25.24  & $-$40.70 &   8.72   &   15.90   & 108.22   & $-$44.85    &        &             &              &     16.9  & 1.04 \\
\multicolumn{12}{c}{\hfill}\\
\bf Model 5     & \bf 25.00  &\bf $-$37.84 &\bf $-$40.00 &  \bf   24.00 &  \bf   5.05 & \bf  $-$40.00  &        &             &              &  \bf  $-1$6.8 &\bf  1.01 \\
FS-JF      &  38.70  & $-$13.65 &    39.62 &           &          &             &        &             &              &    $-1$3.7 &  1.25 \\
FS-KK      &  47.20  & $-$13.80 &    40.50 &           &          &             &        &             &              &    $-1$3.8 &  1.42 \\
FS-LR      &  28.00  & $-$13.75 &    39.50 &           &          &             &   6.00 &  $-$19.00   &   92.00      &    $-1$3.8 &  1.60 \\
FS-MB      &  39.32  & $-$13.5  &    39.3  &           &          &             &        &             &              &    $-1$3.5 &  1.25 \\
FS-MBn     &  39.45  & $-$13.2  &    38.4  &           &          &             &        &             &              &    $-1$3.2 &  1.25 \\
FS-MW      &  35.60  & $-$14.50 &$-$128.00 &           &          &             &        &             &              &    $-1$4.5 &  1.68 \\
FS-RvW     &         &          &          &           &          &             &        &             &              &            &       \\
Wavelet-RS &         &          &          &           &          &             &  17.00 &  $-$14.00   &   97.00      &    $-1$4.0 &  5.20 \\
CS-JS      &  29.10  & $-$13.40 &    39.00 &      3.70 &    34.30 &    34.40    &        &             &              &     $-$8.0 &  1.38 \\
CS-XS      &  23.45  &  $-$5.72 &     0.00 &     11.05 & $-$51.52 &     0.00    &        &             &              &    $-2$0.4 &  1.50 \\
QU-AS      &  29.7   & $-$37.57 & $-$13.21 &     49.52 & $-$14.16 &  $-$155.45  &        &             &              &    $-2$2.9 &  1.00 \\
QU-SO'S    &  24.00  & $-$37.0  & $-$40.4  &     28.00 &     3.4  &   $-$34.6   &        &             &              &    $-1$5.2 &  0.99 \\
QU-TO'B    &  25.07  & $-$36.50 & $-$39.88 &     30.04 &     2.30 &   $-$32.23  &        &             &              &    $-1$5.3 &  1.00 \\
\multicolumn{12}{c}{\hfill}\\
\bf Model 6     & \bf 25.00  &\bf $-$37.84 & \bf  $-$40.00&   \bf  9.00 &   \bf  5.05 &\bf $-$40.00    &        &             &              & \bf   $-$26.5 & \bf 1.00 \\
FS-JF      &  27.38  & $-$31.25 &   $-$69.52&          &          &             &        &             &              &    $-$31.3 &  1.06 \\
FS-KK      &  32.90  & $-$30.80 &   $-$70.50&          &          &             &        &             &              &    $-$30.8 &  1.13 \\
FS-LR      &  22.00  & $-$31.50 &   $-$68.00&          &          &             &   3.00 &  $-$16.00   &  102.00      &    $-$31.5 &  1.16 \\
FS-MB      &  27.81  & $-$33.1  &   $-$63.6 &          &          &             &        &             &              &    $-$33.1 &  1.06 \\
FS-MBn     &  27.81  & $-$31.9  &   $-$67.7 &          &          &             &        &             &              &    $-$31.9 &  1.06 \\
FS-MW      &  26.90  & $-$34.30 &      42.00&     2.00 &$-$648.00 & $-$27.00    &        &             &              &    $-$76.8 &  8.74 \\
FS-RvW     &         & $-$50.00 &           &          &          &             &        &             &              &    $-$50.0 &       \\ 
Wavelet-RS &  12.00  & $-$31.00 &   $-$71.00&     4.00 & $-$70.00 & $-$84.00    &        &             &              &    $-$40.8 &  1.62 \\
CS-JS      &  22.70  & $-$31.00 &   $-$69.80&          &          &             &        &             &              &    $-$31.0 &  1.13 \\
CS-XS      &  11.03  & $-$40.07 &   $-$45.00&     7.61 & $-$51.52 &     0.00    &        &             &              &    $-$44.7 &  1.38 \\
QU-AS      &  27.1   & $-$35.17 &   $-$48.75&     9.17 &    12.83 & $-$76.14    &        &             &              &    $-$23.0 &  1.00 \\
QU-SO'S    &  26.00  & $-$38.0  &   $-$40.8 &     8.00 &    12.9  & $-$68.3     &        &             &              &    $-$26.0 &  1.00 \\
QU-TO'B    &  25.43  & $-$38.15 &   $-$38.41&     9.47 &     7.88 & $-$52.57    &        &             &              &    $-$25.7 &  1.00 \\
\multicolumn{12}{c}{\hfill}\\
\bf Model 7     &\bf  25.00  &\bf $-$44.55 &  \bf   0.00  &  \bf  16.70 &   \bf  37.50 &  \bf   72.00  &        &             &              &  \bf  $-$11.7 &\bf  1.01 \\
FS-JF      &  32.41  & $-$17.55 &    86.09  &          &           &            &        &             &              &    $-$17.6 &  1.13 \\
FS-KK      &  37.80  & $-$16.40 &    82.50  &          &           &            &        &             &              &    $-$16.4 &  1.21 \\
FS-LR      &  28.00  & $-$17.50 &    85.00  &          &           &            &        &             &              &    $-$17.5 &  1.19 \\
FS-MB      &  33.65  & $-$18.8  &    90.2   &          &           &            &        &             &              &    $-$18.8 &  1.14 \\
FS-MBn     &  33.52  & $-$18.3  &    88.4   &          &           &            &        &             &              &    $-$18.3 &  1.14 \\
FS-MW      &  30.70  & $-$22.60 &    19.00  &          &           &            &        &             &              &    $-$22.6 & 11.62 \\
FS-RvW     &         & $-$10.00 &           &          &           &            &        &             &              &    $-$10.0 &       \\  
Wavelet-RS &  15.00  & $-$16.00 &    82.00  &          &           &            &        &             &              &    $-$16.0 &  1.94 \\
CS-JS      &  25.50  & $-$18.00 &    86.50  &          &           &            &        &             &              &    $-$18.0 &  1.23 \\
CS-XS      &  23.21  &  $-$7.63 &    45.00  &     6.92 &  $-$72.51 &     90.00  &        &             &              &    $-$22.5 &  1.27 \\
QU-AS      &  31.04  & $-$29.86 & $-$51.04  &     9.60 &     66.48 &    147.90  &        &             &              &     $-$7.1 &  1.01 \\
QU-SO'S    &  28.00  & $-$38.0  & $-$22.0   &    13.00 &     46.2  &     40.6   &        &             &              &    $-$11.3 &  1.00 \\
QU-TO'B    &  28.28  & $-$36.99 & $-$25.66  &    12.33 &     47.23 &     36.69  &        &             &              &    $-$11.4 &  1.00 \\
\multicolumn{12}{c}{\hfill}\\
\bf Model 8      & \bf 25.00  &\bf 232.56   &\bf  40.00    & \bf  9.00    &\bf 192.70   & \bf  40.00   &        &             &              &   \bf  222.0 &\bf  0.97 \\
FS-JF      &  26.04  & 230.76   &  55.23    &           &          &           &        &             &              &     230.8 &  1.06 \\
FS-KK      &  31.20  & 230.30   &  56.50    &           &          &           &        &             &              &     230.3 &  1.12 \\
FS-LR      &  19.00  & 231.00   &  55.00    &           &          &           &        &             &              &     231.0 &  1.19 \\
FS-MB      &  26.10  & 231.0    &  54.7     &           &          &           &        &             &              &     231.0 &  1.06 \\
FS-MBn     &  26.37  & 231.3    &  53.6     &           &          &           &        &             &              &     231.3 &  1.06 \\
FS-MW      &  23.30  & 229.50   &  95.00    &   2.00   &$-$649.00  &   36.00   &        &             &              &     160.1 &  3.36 \\
FS-RvW     &         &          &           &           &          &           &        &    200.00   &              &     200.0 &       \\
Wavelet-RS &  11.00  & 230.00   &  58.00    &           &          &           &        &             &              &     230.0 &  1.66 \\
CS-JS      &  19.90  & 231.00   &  55.00    &           &          &           &        &             &              &     231.0 &  1.16 \\
CS-XS      &   9.04  & 230.89   &  45.00    &   7.90    & 255.69   &$-$45.00   &        &             &              &     242.5 &  1.16 \\
QU-AS      &  22.00  & 242.18   &$-$173.22  &  10.85    & 192.45   &   28.51   &        &             &              &     225.7 &  0.96 \\
QU-SO'S    &  22.00  & 243.6    &   3.4     &  11.00    & 190.6    &   34.3    &        &             &              &     225.9 &  0.96 \\
QU-TO'B    &  26.33  & 231.47   &  52.73    &   3.47    &  10.40   &   38.94   &        &             &              &     205.7 &  1.03 \\
\multicolumn{12}{c}{\hfill}\\
\bf Model 9     & \bf 25.00  &\bf $-$37.83 & \bf  $-$40.00&   \bf   16.50 &  \bf    5.05 &  \bf  140.00  &          &             &            &    \bf   $-$20.8 &\bf  0.99 \\
FS-JF      &  31.61  & $-$21.85 &  $-$106.88&            &           &            &          &             &            &       $-$21.9 &  1.19 \\
FS-KK      &  38.80  & $-$21.10 &  $-$109.00&            &           &            &          &             &            &       $-$21.1 &  1.31 \\
FS-LR      &  24.00  & $-$22.00 &  $-$106.50&            &           &            &          &             &            &       $-$22.0 &  1.36 \\
FS-MB      &  32.19  & $-$22.6  &  $-$104.3 &            &           &            &          &             &            &       $-$22.6 &  1.19 \\
FS-MBn     &  32.47  & $-$22.5  &  $-$104.3 &            &           &            &          &             &            &       $-$22.5 &  1.19 \\
FS-MW      &  29.30  & $-$28.50 &      46.70&            &           &            &          &             &            &       $-$28.5 &  7.07 \\
FS-RvW     &         & $-$25.00 &       0.00&            &           &            &          &             &            &       $-$25.0 &       \\
Wavelet-RS &  14.00  & $-$21.00 &  $-$110.00&            &           &            &          &             &            &       $-$21.0 &  2.06 \\
CS-JS      &  23.40  & $-$21.90 &  $-$107.30&       3.84 &  $-$57.80 &     32.60  &          &             &            &       $-$27.0 &  1.26 \\
CS-XS      &         &          &           &            &           &            &     6.00 &  $-$22.00   &   56.00    &       $-$22.0 &  1.31 \\
QU-AS      &  34.50  & $-$24.82 &      98.03&      19.56 &      7.89 &  $-$67.95  &          &             &            &       $-$13.0 &  0.98 \\
QU-SO'S    &  30.00  & $-$29.1  &   $-$69.7 &      16.00 &     10.7  &  $-$70.9   &          &             &            &       $-$15.2 &  0.98 \\
QU-TO'B    &  31.62  & $-$28.35 &   $-$70.22&      18.46 &      8.27 &  $-$64.39  &          &             &            &       $-$14.8 &  0.98 \\
\multicolumn{12}{c}{\hfill}\\
\bf Model 10      & \bf 25.00  &\bf $-$37.84 & \bf    0.00  &  \bf   9.00 & \bf  103.00 &\bf $-$36.00 &          &            &         &         \bf   $-$0.6 &\bf  1.08 \\
FS-JF      &  26.35  & $-$40.25 &     8.07  &    14.78 &   115.25 & $-$77.72 &          &            &         &              15.6 &  1.19 \\
FS-KK      &  30.00  & $-$41.10 &    10.50  &     9.65 &   109.50 &   121.50 &          &            &         &            $-$4.4 &  1.17 \\
FS-LR      &  23.00  & $-$39.00 &     6.00  &     6.50 &   112.00 & $-$65.00 &          &            &         &            $-$5.7 &  1.12 \\
FS-MB      &  23.63  & $-$41.6  &    12.8   &     0.60 &   120.1  &    82.8  &          &            &         &           $-$37.6 &  1.27 \\
FS-MBn     &  23.90  & $-$41.7  &    13.4   &     0.61 &   125.0  &    68.5  &          &            &         &           $-$37.6 &  1.27 \\
FS-MW      &  22.50  & $-$45.00 &  $-$4.00  &     7.70 &   103.00 &    89.00 &          &            &         &            $-$7.3 &  3.17 \\
FS-RvW     &         & $-$50.00 &           &          &          &          &          & $-$100.00  &         &                   &       \\
Wavelet-RS &  14.00  & $-$41.00 &    11.00  &     3.00 &   113.00 & $-$69.00 &          &            &         &           $-$13.8 &  1.54 \\
CS-JS      &  23.20  & $-$40.20 &     7.20  &     6.08 &   112.10 &    23.70 &          &            &         &            $-$8.6 &  1.71 \\
CS-XS      &  15.92  & $-$40.07 &     0.00  &     4.83 & $-$38.16 &  $-$0.01 &          &            &         &           $-$39.6 &  1.44 \\
QU-AS      &  24.18  & $-$37.73 &   179.57  &     9.50 &   105.66 &   131.91 &          &            &         &               2.7 &  1.08 \\
QU-SO'S    &  24.00  & $-$37.0  &  $-$3.0   &     9.00 &   102.4  & $-$36.2  &          &            &         &               1.0 &  0.98 \\
QU-TO'B    &  24.49  & $-$36.59 &  $-$4.34  &     9.20 &   100.81 & $-$30.84 &          &            &         &               0.9 &  1.07 \\
\multicolumn{12}{c}{\hfill}\\
\bf Model 11      &\bf  25.00  &\bf 149.50   &\bf  40.00    &\bf  23.75  & \bf  163.50  &\bf $-$68.00   &          &            &         &       \bf   156.3 & \bf 1.16 \\
FS-JF      &  21.66  & 153.86   &$-$2.09    &         &           &            &          &            &         &          153.9 &  1.17 \\
FS-KK      &  25.70  & 154.20   &$-$3.00    &         &           &            &          &            &         &          154.2 &  1.21 \\
FS-LR      &  17.00  & 153.75   &$-$2.00    &         &           &            &          &            &         &          153.8 &  1.23 \\
FS-MB      &  21.84  & 152.6    &   2.3     &         &           &            &          &            &         &          152.6 &  1.17 \\
FS-MBn     &  21.84  & 153.6    &$-$1.3     &         &           &            &          &            &         &          153.6 &  1.17 \\
FS-MW      &  21.20  & 144.50   &  57.80    &   2.0   &$-$800.00  &$-$155.60   &          &            &         &           63.1 &  2.30 \\
FS-RvW     &         & 150.00   &           &         &   250.00  &            &          &            &         &                &       \\
Wavelet-RS &  10.00  & 154.00   &$-$3.00    &         &           &            &          &            &         &          154.0 &  1.53 \\
CS-JS      &  18.20  & 153.20   &   0.00    &         &           &            &          &            &         &          153.2 &  1.20 \\
CS-XS      &  13.74  & 139.29   &  45.00    &   6.38  &   187.00  &$-$135.00   &          &            &         &          154.4 &  1.24 \\
QU-AS      &  21.70  & 153.40   & 178.74    &         &           &            &          &            &         &          153.4 &  1.17 \\
QU-SO'S    &  17.00  & 143.1    &  46.9     &  11.00  &   175.8   & $-$98.5    &          &            &         &          155.9 &  1.14 \\
QU-TO'B    &  17.17  & 143.35   &  45.27    &  10.3   &   177.50  &$-$105.22   &          &            &         &          155.9 &  1.14 \\
\multicolumn{12}{c}{\hfill}\\
\bf Model 12      &\bf  25.00  &\bf $-$232.56 &   \bf  0.00  &  \bf   9.00&  \bf  $-$50.10 & \bf   72.00 &          &            &         &       \bf   $-$184.3 & \bf 1.02 \\
FS-JF      &  25.19  &$-$235.56 &     9.66  &         &             &          &          &            &         &          $-$235.6 &  1.26 \\
FS-KK      &  28.60  &$-$234.00 &     4.00  &    11.40&    $-$58.30 &    99.50 &          &            &         &          $-$183.9 &  1.03 \\
FS-LR      &  19.00  &$-$236.00 &     8.50  &     3.90&    $-$51.50 &    73.00 &          &            &         &          $-$204.6 &  1.24 \\
FS-MB      &  26.65  &$-$230.4  &  $-$9.6   &     0.76&    $-$72.4  &   149.9  &          &            &         &          $-$226.0 &  1.27 \\
FS-MBn     &  26.24  &$-$230.7  &  $-$8.2   &     0.71&    $-$81.2  &     0.7  &          &            &         &          $-$226.8 &  1.26 \\
FS-MW      &  25.50  &$-$237.20 & $-$54.00  &     4.00&    $-$43.30 &   158.50 &          &            &         &          $-$210.9 &  7.44 \\
FS-RvW     &         &$-$200.00 &           &         &        0 00 &          &          &            &         &                   &       \\
Wavelet-RS &  13.00  &$-$233.00 &     0.00  &     4.00&    $-$74.00 & $-$28.00 &          &            &         &          $-$195.6 &  1.56 \\
CS-JS      &  16.50  &$-$235.60 &    16.50  &         &             &          &          &            &         &          $-$235.6 &  1.51 \\
CS-XS      &  22.83  &$-$234.70 &     0.00  &     6.05&    $-$57.24 &    90.00 &          &            &         &          $-$197.5 &  1.17 \\
QU-AS      &  26.80  &$-$231.82 &  $-$2.98  &     9.89&    $-$61.57 & $-$68.58 &          &            &         &          $-$185.9 &  1.01 \\
QU-SO'S    &  27.00  &$-$231.6  &  $-$5.5   &    10.00&    $-$58.7  &   100.5  &          &            &         &          $-$184.8 &  1.01 \\
QU-TO'B    &  26.66  &$-$231.52 &  $-$5.60  &    10.21&    $-$58.74 &   100.64 &          &            &         &          $-$183.7 &  1.01 \\
\multicolumn{12}{c}{\hfill}\\
\bf Model 13      & \bf 25.00  &\bf $-$44.55 &   \bf  0.00  &  \bf  24.00 &   \bf  37.54 &   \bf  72.00 &          &            &         &          \bf   $-$4.3 &\bf  0.99 \\
FS-JF      &  37.88  &  $-$4.65 &    39.67  &          &           &           &          &            &         &             $-$4.7 &  1.22 \\
FS-KK      &  44.50  &  $-$4.80 &    40.00  &          &           &           &          &            &         &             $-$4.8 &  1.33 \\
FS-LR      &  26.00  &  $-$2.00 &    78.00  &          &           &           &          &            &         &             $-$2.0 &  7.27 \\
FS-MB      &  25.82  &  $-$5.4  &    42.6   &          &           &           &          &            &         &             $-$5.4 &  1.61 \\
FS-MBn     &  25.69  &  $-$4.9  &    40.5   &          &           &           &          &            &         &             $-$4.9 &  1.61 \\
FS-MW      &  34.60  & $-$13.00 &    21.20  &          &           &           &          &            &         &            $-$13.0 &  8.70 \\
FS-RvW     &         &     0.00 &           &          &           &           &          &            &         &                0.0 &       \\
Wavelet-RS &  17.00  &  $-$5.00 &    40.00  &          &           &           &          &            &         &             $-$5.0 &  2.38 \\
CS-JS      &  28.80  &  $-$4.90 &    40.40  &          &           &           &          &            &         &             $-$4.9 &  1.44 \\
CS-XS      &  13.81  & $-$34.35 & $-$45.00  &    12.83 &     30.53 &     90.00 &          &            &         &             $-$3.1 &  1.25 \\
QU-AS      &  31.7   & $-$28.01 &   121.79  &    16.62 &     55.69 &      7.18 &          &            &         &                0.8 &  0.99 \\
QU-SO'S    &  27.00  & $-$38.5  & $-$21.1   &    22.00 &     41.9  &     55.9  &          &            &         &             $-$2.4 &  0.98 \\
QU-TO'B    &  27.08  & $-$38.55 & $-$20.98  &    21.72 &     41.89 &     55.90 &          &            &         &             $-$2.7 &  0.98 \\
\tableline
\multicolumn{12}{c}{\hfill}\\
\multicolumn{12}{c}{\normalsize \em Faraday thick component}\\
\multicolumn{12}{c}{\hfill}\\
\tableline
\bf Model 14      &         &          &           &         &            &           &  \bf 1.90 &\bf $-$136.98  & \bf  50.00  &   \bf  $-$137.0  &\bf 0.97 \\
FS-JF      &   6.03  & $-$71.95 &    34.49  &     5.71& $-$210.14  &   70.24   &        &           &          &     $-$139.2  & 1.09 \\
FS-KK      &  16.10  & $-$68.90 &    24.50  &    12.70& $-$192.80  &   18.00   &        &           &          &     $-$123.5  & 4.80 \\
FS-LR      &         &          &           &         &            &           &   1.00 &$-$117.50  &  305.00  &     $-$117.0  & 2.11 \\
FS-MB      &   4.37  &$-$193.1  &    16.0   &     0.92&  $-$80.6   &   58.6    &        &           &          &     $-$173.5  & 1.40 \\
FS-MBn     &   4.47  &$-$201.7  &    44.6   &     0.91&  $-$79.6   &   54.9    &        &           &          &     $-$181.0  & 1.37 \\
FS-MW      &   4.00  &$-$211.80 & $-$21.50  &     5.80&  $-$83.00  &   82.00   &        &           &          &     $-$135.6  & 2.56 \\
FS-RvW     &         & $-$75.00 &           &         &    175.00  &           &        &           &          &               &      \\
Wavelet-RS &   7.00  & $-$72.00 &    34.00  &         &            &           &   6.00 &$-$140.00  &   69.00  &     $-$103.0  & 2.01 \\
CS-JS      &   1.25  & $-$72.20 &    37.80  &     0.26& $-$208.10  &   63.90   &        &           &          &      $-$95.6  & 1.80 \\
CS-XS      &   3.99  & $-$74.42 &    45.00  &         &            &           &        &           &          &      $-$74.4  & 1.05 \\
QU-AS      &   8.63  &$-$165.85 & $-$82.75  &     8.90& $-$114.95  &    2.33   &        &           &          &     $-$139.9  & 0.93 \\
QU-SO'S    &   6.00  &$-$206.5  &    51.0   &         &            &           &  57.00 &           &70.8   &     $-$206.5  & 1.5  \\
QU-TO'B    &   0.89  &$-$190.70 &  $-$0.09  &         &            &           &  19.47 &$-$101.40  &   40.00  &     $-$101.0  & 1.13 \\
\multicolumn{12}{c}{\hfill}\\
\bf Model 15      & \bf  1.80  &\bf $-$240.22  & \bf $-$36.00  &        &               &         &    \bf  1.90  &\bf $-$250.17  &  \bf  50.00  &   \bf    $-$245.0 &\bf  1.03 \\
FS-JF      &   7.29  &$-$241.56  &  $-$31.09  &        &               &         &            &            &           &       $-$241.6 &  1.06 \\
FS-KK      &  31.30  &$-$241.30  &  $-$31.50  &        &               &         &            &            &           &       $-$241.3 & 23.87 \\
FS-LR      &   5.90  &$-$241.50  &  $-$30.50  &        &               &         &            &            &           &       $-$241.5 &  1.13 \\
FS-MB      &   7.19  &$-$241.9   &  $-$30.1   &        &               &         &            &            &           &       $-$241.9 &  1.06 \\
FS-MBn     &   7.19  &$-$241.2   &  $-$32.5   &        &               &         &            &            &           &       $-$241.2 &  1.06 \\
FS-MW      &   6.60  &$-$259.50  &  $ $10.00  &        &               &         &            &            &           &       $-$259.5 &  2.02 \\
FS-RvW     &         &$-$250.00  &            &        &      150.00   &         &            &            &           &                &       \\
Wavelet-RS &  17.00  &$-$241.00  &  $-$32.00  &        &               &         &            &            &           &       $-$241.0 &  4.80 \\
CS-JS      &         &           &            &        &               &         &      10.53 & $-$241.80  &    24.80  &       $-$242.0 &  1.03 \\
CS-XS      &   3.38  &$-$253.78  &  $ $ 0.00  &        &               &         &            &            &           &       $-$253.8 &  1.18 \\
QU-AS      &   7.26  &$-$241.30  &  $1$49.13  &        &               &         &            &            &           &       $-$241.3 &  1.05 \\
QU-SO'S    &   7.00  &$-$241.6   &  $-$30.9   &        &               &         &            &            &           &       $-$241.6 &  1.06 \\
QU-TO'B    &         &           &            &        &               &         &      10.92 & $-$241.62  &    25.40  &       $-$252.0 &  1.03 \\
\multicolumn{12}{c}{\hfill}\\
\bf Model 16      &         &           &            &        &          &              &  \bf  1.90 &\bf $-$136.98  & \bf   25.00   &      \bf $-$137.0 &       \\  
FS-JF      &   6.67  &$-$139.46  &   46.91    &        &          &              &         &            &            &       $-$139.5 &  1.03 \\
FS-KK      &  19.40  &$-$139.20  &   46.00    &        &          &              &         &            &            &       $-$139.2 &  3.96 \\
FS-LR      &   5.50  &$-$138.75  &   46.00    &        &          &              &         &            &            &       $-$138.8 &  1.05 \\
FS-MB      &   6.63  &$-$139.8   &   48.2     &        &          &              &         &            &            &       $-$139.8 &  1.03 \\
FS-MBn     &   6.63  &$-$139.8   &   48.0     &        &          &              &         &            &            &       $-$139.8 &  1.03 \\
FS-MW      &   7.40  &$-$152.50  &$-$60.00    &        &          &              &         &            &            &       $-$152.5 &  1.94 \\
FS-RvW     &         &$-$150.00  &            &        &  $-$50.00&              &         & $-$250.00  &            &                &       \\
Wavelet-RS &  10.00  &$-$139.00  &   46.00    &        &          &              &         &            &            &       $-$139.0 &  1.23 \\
CS-JS      &   5.65  &$-$138.90  &   45.30    &        &          &              &         &            &            &       $-$138.9 &  1.05 \\
CS-XS      &   1.99  &$-$141.20  &   45.00    &   1.97 & $-$156.47&   $-$90.00   &         &            &            &       $-$148.8 &  1.25 \\
QU-AS      &   6.68  &$-$138.75  &   45.25    &        &          &              &         &            &            &       $-$138.8 &  1.03 \\
QU-SO'S    &   7.00  &$-$139.3   &   46.2     &        &          &              &         &            &            &       $-$139.3 &  1.03 \\
QU-TO'B    &   5.71  &$-$144.60  &   45.00    &   4.23 & $-$128.27&      38.44   &         &            &            &       $-$137.6 &  1.02 \\
\multicolumn{12}{c}{\hfill}\\
\bf Model 17      & \bf  1.80  &\bf $-$240.00  &\bf $-$36.00    &        &          &              & \bf   1.90 &\bf $-$250.17  & \bf   25.00   &    \bf   $-$245.2 &       \\ 
FS-JF      &  10.78  &$-$245.36  &$-$18.29    &        &          &              &         &            &            &       $-$245.4 &  0.97 \\
FS-KK      &  47.40  &$-$244.80  &    7.50    &        &          &              &         &            &            &       $-$244.8 & 53.95 \\
FS-LR      &  10.20  &$-$245.00  &$-$18.25    &        &          &              &         &            &            &       $-$245.0 &  0.98 \\
FS-MB      &   3.02  &$-$245.4   &$-$19.6     &        &          &              &         &            &            &       $-$245.4 &  0.97 \\
FS-MBn     &   3.02  &$-$245.0   &$-$18.4     &        &          &              &         &            &            &       $-$245.0 &  0.97 \\
FS-MW      &   9.90  &$-$247.50  &$-$18.50    &        &          &              &         &            &            &       $-$247.5 &  2.52 \\
FS-RvW     &         & $-$17.80  &            &        &          &              &         & $-$250.00  &            &                &       \\
Wavelet-RS &  22.00  &$-$245.00  &$-$18.40    &        &          &              &         &            &            &       $-$245.0 &  5.97 \\
CS-JS      &  10.40  &$-$245.40  &$-$20.00    &        &          &              &         &            &            &       $-$245.4 &  0.97 \\
CS-XS      &   5.47  &$-$238.52  &$-$20.00    &        &          &              &         &            &            &       $-$238.5 &  1.20 \\
QU-AS      &  10.75  &$-$245.11  &$-$18.05    &        &          &              &         &            &            &       $-$245.1 &  0.96 \\
QU-SO'S    &  11.00  &$-$245.3   &$-$18.4     &        &          &              &         &            &            &       $-$245.3 &  0.97 \\  
QU-TO'B    &  10.79  &$-$245.33  &    0.00    &        &          &              &         &            &            &       $-$245.3 &  0.97 \\
\enddata
\end{deluxetable*}
\end{document}